\documentclass[11pt]{article}

\usepackage[a4paper,left=20mm,top=20mm,right=20mm,bottom=25mm]{geometry}

\usepackage{amsmath, amsfonts, amssymb, amsthm}
\usepackage{mathtools}
\usepackage[mathscr]{euscript}
\usepackage{wasysym}

\usepackage{graphicx}
\usepackage{xcolor}
\usepackage{enumitem}
\usepackage{authblk}

\usepackage{color}
\usepackage[colorlinks=true,urlcolor=black,linkcolor=blue,citecolor=blue]{hyperref}
\usepackage[font=footnotesize,width=.85\textwidth,labelfont=bf]{caption}

\usepackage[numbers,sort]{natbib}

\newtheorem{theorem}{Theorem}[section]
\newtheorem{lemma}[theorem]{Lemma}
\newtheorem{definition}[theorem]{Definition}
\newtheorem{corollary}[theorem]{Corollary}
\newtheorem*{problem}{Problem}
\newtheorem{proposition}[theorem]{Proposition}
\newtheorem{fact}[theorem]{Observation}
\newtheorem*{remark}{Remark}

\DeclareMathOperator{\cl}{cl}
\DeclareMathOperator{\lca}{lca}
\DeclareMathOperator{\child}{child}
\DeclareMathOperator{\parent}{par}
\DeclareMathOperator{\symdiff}{\triangle}
\DeclareMathOperator{\res}{res}

\newcommand{\eps}{\ensuremath{\varepsilon}}
\newcommand{\irr}[1]{\ensuremath{[#1]_{\textrm{irr}}}}
\newcommand{\eXM}{\ensuremath{\eps \colon \irr{X \times X} \to 2^{M}}}

\newcommand{\HP}{\mathcal{H}^*_{\mathcal{P}}}

\makeatletter
\def\moverlay{\mathpalette\mov@rlay}
\def\mov@rlay#1#2{\leavevmode\vtop{%
    \baselineskip\z@skip \lineskiplimit-\maxdimen
    \ialign{\hfil$\m@th#1##$\hfil\cr#2\crcr}}}
\newcommand{\charfusion}[3][\mathord]{
  #1{\ifx#1\mathop\vphantom{#2}\fi
    \mathpalette\mov@rlay{#2\cr#3}
  }
  \ifx#1\mathop\expandafter\displaylimits\fi}
\DeclareRobustCommand\bigop[1]{%
  \mathop{\vphantom{\sum}\mathpalette\bigop@{#1}}\slimits@
}
\newcommand{\bigop@}[2]{%
  \vcenter{%
    \sbox\z@{$#1\sum$}%
    \hbox{\resizebox{\ifx#1\displaystyle.9\fi\dimexpr\ht\z@+\dp\z@}{!}{$\m@th#2$}}%
  }%
}
\makeatother

\newcommand{\cupdot}{\charfusion[\mathbin]{\cup}{\cdot}}

\newcommand{\AX}[1]{\upshape{#1}}
\newcommand{\mc}{\mathcal}

\providecommand{\keywords}[1]{\textbf{\textit{Keywords: }} #1}

\title{Compatibility of Partitions with Trees, Hierarchies, and Split Systems}

\author[a]{Marc Hellmuth}
\author[b,c]{David Schaller}
\author[b,c,d,e,f]{Peter F.\ Stadler}

\affil[a]{Department of Mathematics, Faculty of Science, Stockholm
  University, SE - 106 91 Stockholm, Sweden
  \authorcr \texttt{marc.hellmuth@math.su.se}}

\affil[b]{Bioinformatics Group, Department of Computer Science \&
  Interdisciplinary Center for Bioinformatics, Leipzig University,
  H{\"a}rtelstra{\ss}e 16–18, D-04107 Leipzig, Germany
  \authorcr \texttt{sdavid@bioinf.uni-leipzig.de} $\cdot$ 
  \texttt{studla@bioinf.uni-leipzig.de}}

\affil[c]{Max Planck Institute for Mathematics in the Sciences,
  Inselstra{\ss}e 22, D-04103 Leipzig, Germany}

\affil[d]{Department of Theoretical Chemistry University of Vienna,
  W{\"a}hringerstra{\ss}e 17, A-1090 Wien, Austria}

\affil[e]{Facultad de Ciencias, Universidad National de Colombia, Sede
  Bogot{\'a}, Colombia}

\affil[f]{Santa Fe Institute, 1399 Hyde Park Rd., Santa Fe, NM 87501, USA}

\date{\ }

\begin{document}
  
\maketitle 

\abstract{
  The question whether a partition $\mathcal{P}$ and a hierarchy
  $\mathcal{H}$ or a tree-like split system $\mathfrak{S}$ are compatible
  naturally arises in a wide range of classification problems. In the
  setting of phylogenetic trees, one asks whether the sets of $\mathcal{P}$
  coincide with leaf sets of connected components obtained by deleting some
  edges from the tree $T$ that represents $\mathcal{H}$ or $\mathfrak{S}$,
  respectively.  More generally, we ask whether a refinement $T^*$ of $T$
  exists such that $T^*$ and $\mathcal{P}$ are compatible in this
  sense.  The latter is closely related to the question as to whether
  there exists a tree at all that is compatible with $\mathcal{P}$.  We
  report several characterizations for (refinements of) hierarchies and
  split systems that are compatible with (systems of) partitions. In
  addition, we provide a linear-time algorithm to check whether refinements
  of trees and a given partition are compatible. The latter problem becomes
  NP-complete but fixed-parameter tractable if a system of partitions
  is considered instead of a single partition. In this context, we
  also explore the close relationship of the concept of compatibility
  and so-called Fitch maps.
}

\bigskip
\noindent

\sloppy

\section{Introduction}

The selection of a partition $\mathcal{P}$ from a hierarchy $\mathcal{H}$
(or its equivalent rooted tree $T$) on a finite set $X$ is often the
final step in applications of hierarchical clustering procedures
\cite{Milligan:85,VegaPons:10}.  In this case, $\mathcal{P}$, often called
a ``representative partition'', is composed of the pairwise disjoint leaf
sets $L(T(u))$ of subtrees $T(u)$ rooted at some vertices $u$ of
$T$. Equivalently, each set (class) of $\mathcal{P}$ is a set in
$\mathcal{H}$, i.e., $\mathcal{P}\subseteq\mathcal{H}$.  Cutting a
  hierarchy at different ``levels'' leads to partitions $\mathcal{P}_i$ of
  $X$ that are ordered by refinement, i.e., every set $A\in\mathcal{P}_i$
  at the lower level is contained in a set $B\in\mathcal{P}_j$ at the
  higher level. In the image processing literature, braids of partitions
have been introduced as systems $\{\mathcal{P}_i\mid i=1,\dots, k\}$
of partitions of $X$ that generalize such hierarchically ordered
  partitions. In a braid, all pairwise refinement suprema
  $\mathcal{P}_i\vee \mathcal{P}_{j\ne i}$ (i.e., the finest partition that
  is refined by both $\mathcal{P}_i$ and $\mathcal{P}_{j}$) must be
  hierarchically organized w.r.t.\ refinement, and moreover satisfy
  $\{X\}\ne \mathcal{P}_i\vee \mathcal{P}_j$
\cite{Kiran:15,Tochon:19}. Considering the distribution of attributes of
species, biologists have been interested in systems of partitions
$\{\mathcal{P}_1,\dots,\mathcal{P}_k\}$ and the associated system of splits
$A|(X\setminus A)$ with $A\in\mathcal{P}_i$ for some $i$. In
\cite{Huber:14}, the compatible split systems that are generated by
partition systems are characterized.

Instead of considering systems of partitions, one can ask whether a single
partition $\mathcal{P}$, or -- equivalently -- its associated split system
$\mathfrak{S}_{\mathcal{P}} \coloneqq \left\{A|(X\setminus A)\, \colon
  A\in\mathcal{P}\right\}$ can be obtained from a rooted or unrooted tree
$T$ by cutting a subset $H\subseteq E(T)$ of the tree edges and considering
the leaf sets of the resulting connected components. For rooted trees, this
question arises naturally in mathematical phylogenetics. The removal of
those edges from the rooted gene tree that correspond to horizontal gene
transfer (HGT) then leaves subtrees of the gene phylogeny that can be
analyzed independently \cite{Hellmuth:2017,Geiss:18a,Hellmuth:18a}. If the
tree $T$ and the partition $\mathcal{P}$ of the leaf set into HGT-free
subsets are inferred independently, it is important to
recognize whether the data are compatible with each other, i.e., whether
$\mathcal{P}$ can be obtained from the tree $T$ by cutting some of its
edges. If this is not possible for a tree that contains multifurcations, it
may still be possible to achieve compatibility by refining some of the
multifurcations in $T$. Here, we address these two main questions.

The split system $\mathfrak{S}_{\mathcal{P}}$ introduced above suggests a
different notion of compatibility with trees. Consider the partition
$\mathcal{P}=\{\{a\},\{b\},\{c,d\}\}$. Clearly, $\mathcal{P}$ is compatible
with the star tree $S_4$ on $X=\{a,b,c,d\}$. However, $S_4$ (more
precisely, its unrooted version $\overline{S_4}$) does not display the
split $\{c,d\}\mid(X\setminus\{c,d\})$. In fact, the condition that
$\mathfrak{S}_{\mathcal{P}}$ is displayed by the unrooted tree
$\overline{T}$ is closely related to the idea that $\mathcal{P}$ is a
representative partition (of a suitably rooted version) of the tree
$\overline{T}$. This example shows that the notion of compatibility
considered here is more general than the concepts
that have appeared in the literature so far.

In this contribution, we characterize compatibility of partitions with
  rooted and unrooted trees (or equivalently their hierarchies and split
  systems, respectively). After introducing the notation and some
  preliminary results, we give an overview of the main concepts and results
  in Section~\ref{sec:overview}.

\section{Preliminaries}
\label{sec:prelim}

\paragraph{\textbf{Basics}}
We denote the power set of a set $X$ by $2^X$.  A set system
$\mathcal{P}\subseteq 2^X$ is a \emph{partition} of $X$ if \AX{(P0)}
$\emptyset\notin\mathcal{P}$, \AX{(P1)} $\bigcup_{A\in\mathcal{P}} A = X$,
and \AX{(P2)} if $A,B\in\mathcal{P}$ and $A\cap B\ne\emptyset$ then $A=B$.
We will interchangeably use the equivalent terms: set of partitions,
collection of partitions and partition systems.  Two sets $A$ and $B$
\emph{overlap} if $A\cap B\ne\emptyset$, $A\setminus B\ne\emptyset$, and
$B\setminus A\ne\emptyset$.

In this contribution, we consider both rooted and unrooted
  phylogenetic trees $T$ with vertex set $V(T)$, edge set $E(T)$, and leaf
  set $L(T)=X$.  In this case, we also say that $T$ is a tree on $X$.  A
  \emph{star tree} $T$ is a tree for which $|V(T)\setminus L(T)|=1$.

\paragraph{\textbf{Rooted Trees and Hierarchies}}
A rooted tree $T$ has a distinguished vertex $\rho_T$ called the
  \emph{root} of $T$.  For $u\in V(T)$, we write $\child_{T}(u)$ for the
set of its children, and $\parent_{T}(u)$ for the parent of
$u\ne\rho_T$. In both cases, we may omit the index $T$ whenever there is no
risk of confusion.  The subtree of $T$ rooted at $u$ is denoted by $T(u)$.
Furthermore, we write $\preceq_T$ for the ancestor partial order on $T$,
that is, $u\preceq_T v$ if $v$ lies on the path from $\rho_T$ to $u$.  If
$u\preceq_T v$ or $v\preceq_T u$, then $u$ and $v$ are comparable and,
otherwise, incomparable.  For a nonempty subset of $A\subseteq X$, we
denote by $\lca_T(A)$ the last common ancestor of $A$ in $T$.  A
  rooted tree is \emph{phylogenetic} if all its inner vertices
$V(T)\setminus L(T)$ have at least two children.  Hence, a rooted
phylogenetic tree may contain one vertex with degree $2$, namely the root 
$\rho_T$.

A set system $\mathcal{H}\subseteq 2^X$ is a \emph{hierarchy (on $X$)} if
\AX{(H0)} $\emptyset\notin\mathcal{H}$, \AX{(H1)} $X\in\mathcal{H}$,
\AX{(H2)} $A,B\in\mathcal{H}$ implies $A\cap B\in\{A,B,\emptyset\}$, i.e.,
$A$ and $B$ do not \emph{overlap}, and \AX{(H3)} $\{x\}\in\mathcal{H}$ for
all $x\in X$. For a given non-empty set $A\subseteq X$ and a hierarchy
$\mathcal{H} \subseteq 2^X$, we define $A_{\mathcal{H}}$ as the
inclusion-minimal element in $\mathcal{H}$ that contains $A$.

For a hierarchy $\mathcal{H} \subseteq 2^X$ we can define the
\emph{closure} as the function $\cl_{\mathcal{H}}:2^X\to 2^X$ satisfying
\begin{equation}
  \cl_{\mathcal{H}}(A) \coloneqq \bigcap_{B\in\mathcal{H},\, A\subseteq B} B
\end{equation}
for all subsets $A\subseteq X$. Where there is no danger of confusion, we
will drop the explicit reference to $\mathcal{H}$ and simply write $\cl(A)$
instead of $\cl_{\mathcal{H}}(A)$.

\begin{lemma}
  Let $A\subseteq X$ be non-empty and $\mathcal{H}$ be a hierarchy on $X$.
  Then $\cl(A) = A_{\mathcal{H}}$ for all $A\ne\emptyset$.
  \label{lem:AH}
\end{lemma}
\begin{proof}
  Since $A\subseteq X$, $X\in \mathcal{H}$, and no two elements in
  $\mathcal{H}$ overlap, there is a unique inclusion-minimal element
  $A_{\mathcal{H}}$ in $\mathcal{H}$ that contains $A$, i.e., every
  $A'\in\mathcal{H}$ that contains $A$ also contains $A_{\mathcal{H}}$.
  Thus $\cl(A)=\bigcap\{B\in\mathcal{H}\mid A\subseteq B\}
  =\bigcap\{B\in\mathcal{H}\mid A_{\mathcal{H}}\subseteq B\} =
  A_{\mathcal{H}}$.
\end{proof}

As an immediate consequence, we observe that for a hierarchy
$\mathcal{H}\subseteq 2^X$, $\cl$ satisfies the classical properties of a
closure operator: \AX{(C1)} $A\subseteq \cl(A)$ (\emph{enlarging});
\AX{(C2)} $\cl(\cl(A))=\cl(A)$ (\emph{idempotent}); \AX{(C3)} If
$A\subseteq B$, then $\cl(A)\subseteq \cl(B)$ (\emph{isotone}).  For
$|X|\ge 2$, $\mathcal{H}$ contains at least two distinct singletons $\{x\}$
and $\{y\}$ and thus $\cl(\emptyset)\subseteq
\{x\}\cap\{y\}=\emptyset$. Only in the special case $|X|=1$, i.e.,
$\mathcal{H}=\{\{x\}\}$, we get $\cl(\emptyset)=\{x\}$. This is the
consequence of the usual practice of excluding $\emptyset$ from hierarchies
to have the singleton as inclusion-minimal elements, which in turn is
motivated by the 1-to-1 correspondence between rooted trees and
hierarchies. We note that in the context of abstract convexities, on the
other hand, it is customary to enforce $\emptyset\in\mathcal{H}$ so that
$\mathcal{H}$ is closed under intersection \cite{vandeVel:93}, in
which case $\cl(\emptyset)=\emptyset$ also for $|X|\le 1$. This subtlety is
irrelevant for our discussion, however, since we are not interested in the
trivial cases $|X|\le 1$.

The following result shows that there is a  1-to-1 correspondence between
  hierarchies and rooted
  phylogenetic trees.
  \begin{theorem}[\cite{Semple:03}]\label{thm:1-1}
    Let $\mathcal H$ be a collection of non-empty subsets of $X$.
    Then, $\mathcal{H}$ is a hierarchy on $X$ if and only if
    there is a rooted phylogenetic tree $T$ on $X$ with
    $\mathcal{H} \coloneqq \{L(T(v)) \mid c \in V (T)\}$.
  \end{theorem}
The sets $L(T(u))$ for $u\in V(T)$ or, equivalently, the sets
		of a hierarchy are commonly referred to as \emph{clusters}.
In view of Thm.~\ref{thm:1-1}, every set $A\in\mathcal{H}$ corresponds to
a vertex $u_A \in V(T)$ such that $L(T(u_A))=A$ and $u_A\preceq_T u_B$ is
equivalent to $A\subseteq B$. The hierarchy $\mathcal{H}(T)$ of a
rooted phylogenetic tree $T$ is $\mathcal{H}(T) \coloneqq \{L(T(v))\mid
v\in V(T)\}$.
Moreover, we say that a rooted tree $T$ \emph{corresponds to}
$\mathcal{H}$ if $\mathcal{H}=\mathcal{H}(T)$.
In particular, Thm.~\ref{thm:1-1} ensures that for all hierarchies there is
a corresponding tree and, moreover, that all results established here for
hierarchies do also hold for rooted phylogenetic trees and \emph{vice versa}.
As an immediate consequence, we can express the closure as
\begin{equation}
  \cl_{\mathcal{H}}(A) = L(T(\lca_T(A))).
\end{equation}
We therefore have $\lca_T(A)\preceq_T\lca_T(B)$ if and only if
$\cl(A)\subseteq\cl(B)$ for all $A,B\ne\emptyset$.
Given a set $A\in \mathcal{H}$ with $|A|>1$ (and corresponding vertex $u\in
V(T)$ that satisfies $A=L(T(u))$), we call $B\in \mathcal{H}$ a \emph{child
cluster of $A$} if $B=L(T(v))$ for some $v\in\child_{T}(u)$. Hence, the
child clusters $B$ of $A$ are exactly the inclusion-maximal sets in
$\mathcal{H}$ that satisfy $B\subsetneq A$.
For simplicity of
notation, we will think of the leaves of $T$ in this case as the $x\in X$,
i.e., $L(T)=X$. A phylogenetic tree $T$ is a \emph{refinement} of the
phylogenetic tree $T'$ on the same set $X$ if the corresponding
hierarchy $\mathcal{H}(T)$ is a refinement of $\mathcal{H}(T')$, i.e., if
$\mathcal{H}(T')\subseteq \mathcal{H}(T)$.

\paragraph{\bf Unrooted Trees and Split Systems}
An unrooted tree is \emph{phylogenetic} if all non-leaf vertices have
degree at least three.  A \emph{split} $A_1|A_2 \coloneqq \{A_1,A_2\}$ on
$X$ is a partition of the set $X$ into two disjoint non-empty subsets $A_1$
and $A_2=X\setminus A_1$. In every tree $T$ on $X$, we can associate with
every edge $e\in E(T)$ the split $\mathcal{S}_e = L(T_1)|L(T_2)$ where
$L(T_1)$ and $L(T_2)$ are the leaf sets of the two (not necessarily
phylogenetic) trees $T_1$ and $T_2$, respectively, obtained from $T$ by
deletion of $e$. An unrooted phylogenetic tree $\overline{T}$ is determined
by its split system
$\mathfrak{S}(\overline{T}) = \{\mathcal{S}_e \colon e\in E(\overline T)
\}$. To be more precise, there is a 1-to-1 correspondence between unrooted
phylogenetic trees $\overline{T}$ with leaf set $X$ and split systems
$\mathfrak{S}$ that (i) contain all ``singleton splits''
$\{x\}|(X\setminus\{x\})$ and (ii) are ``compatible'' in the sense that,
for any two splits $A_1|A_2, B_1|B_2\in\mathfrak{S}$, at least one of the
four intersections $A_1\cap B_1$, $A_1\cap B_2$, $A_2\cap B_1$, or
$A_2\cap B_2$ is empty \cite{Buneman:71}.  In this case, there is a unique
(up to isomorphism) tree $\overline T$ with
$\mathfrak{S}(\overline{T}) = \mathfrak{S}$. We call such split systems
\emph{tree-like}.  An unrooted tree $\overline{T}^*$ is a refinement of
$\overline{T}$ if
$\mathfrak{S}(\overline{T})\subseteq \mathfrak{S}(\overline{T}^*)$.

We note that this is a special case of the analogous result for so-called
$X$-trees, see e.g.\ \cite[Prop.~3.5.4]{Semple:03}. In an $X$-tree, a
  set of ``taxa'' $X$ is mapped (not necessarily injectively) to the vertex
  set $V(T)$ of a rooted or unrooted tree $T$ \cite{Semple:03}. The
  \emph{phylogenetic (rooted or unrooted) trees} considered here are a
  slightly less general construction that identifies the taxa set $X$ with
  the leaf set $L(T)$ and insists that distinct taxa are represented by
  distinct vertices in the underlying trees. That is, they are equivalent
  to $X$-trees $T$ for which $X$ is bijectively mapped to $L(T)$.

\begin{remark}
  Throughout this contribution, we assume that $X$ is finite and
  $|X|\ge 2$. Moreover, all rooted and unrooted trees are phylogenetic
  unless explicitly specified otherwise.
\end{remark}

\section{Main Ideas and Results}
\label{sec:overview}

\begin{figure}[t]
  \begin{center}
    \includegraphics[width=0.85\textwidth]{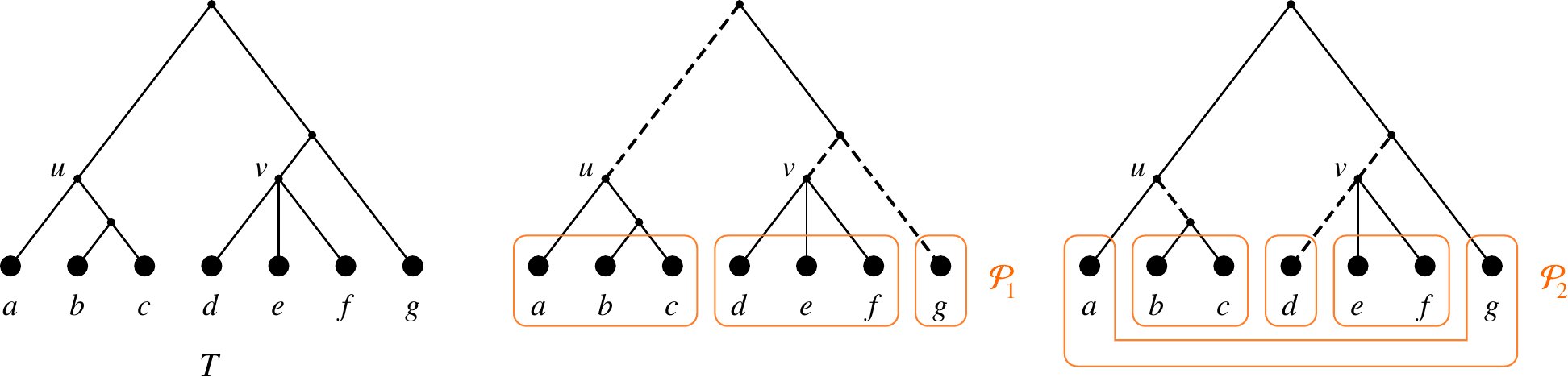}
  \end{center}
  \caption{A tree $T$ on $X=\{a,b,c,d,e,f,g\}$ with two examples for sets
  of separating edges $H_1$ (dashed edges in the middle panel) and $H_2$
  (dashed edges in the right panel). Removal of $H_1$ and $H_2$ induces the
  partitions $\mc{P}_1=\{\{a,b,c\},\{d,e,f\},\{g\}\}$ and
  $\mc{P}_2=\{\{a,g\},\{b,c\},\{d\},\{e,f\}\}$ of $X$, respectively.}
  \label{fig:separating-set-examples}
\end{figure}

Let $T$ be a (rooted or unrooted) tree with leaf set $L(T)=X$ and
$H\subseteq E(T)$ be a subset of edges. Removal of $H$ disconnects $T$ into
a forest whose connected components induce the partition $\mathcal{F}(T,H)$
on the leaf set $X$. Of course, it may be possible that removal of
the edges $H$ separates inner vertices, e.g., if all incident edges to an
inner vertex are in $H$. This, however, does not change the fact that we
still obtain a partition of $X$ after removal of the edges in $H$. We will
refer to the edges $e\in H$ as \emph{separating edges}.
Fig.~\ref{fig:separating-set-examples} shows two examples for sets of
  separating edges, $H_1$ and $H_2$ as indicated by the dashed lines, for a
  given tree $T$.  The partition $\mc{P}_1=\mathcal{F}(T,H_1)$ is a
  ``representative partition'' for $T$, i.e., all sets in $\mc{P}_1$ appear
  as clusters in $T$: $\{a,b,c\}=L(T(u))$, $\{d,e,f\}=L(T(v))$, and
  $\{g\}=L(T(g))$.  In contrast, the partition
  $\mc{P}_2=\mathcal{F}(T,H_2)$ contains the set $\{a,g\}$ which is not a
  cluster in $T$, and thus, $\mc{P}_2$ is not a representative partition
  for $T$.

The notion of compatibility of partitions and trees used here is
  defined in terms of separating edges and the partitions that they
  induce:
\begin{definition}
  Let $\mathcal{P}$ be a partition of $X$ and let $T$ be a rooted or
  unrooted tree with leaf set $X$.  Then $\mathcal{P}$ and $T$ are
  \emph{compatible} if there is a set of separating edges
  $H\subseteq E(T)$ such that $\mathcal{P}=\mathcal{F}(T,H)$.

  In case $T$ is rooted (or unrooted) and compatible with $\mathcal{P}$,
  the corresponding hierarchy $\mathcal{H}(T)$ (or split system
  $\mathfrak{S}(T)$, resp.) are said to be \emph{compatible} with
    $\mathcal{P}$.
  \label{def:compatible-rooted}
\end{definition}

\begin{figure}[t]
  \begin{center}
    \includegraphics[width=.75\textwidth]{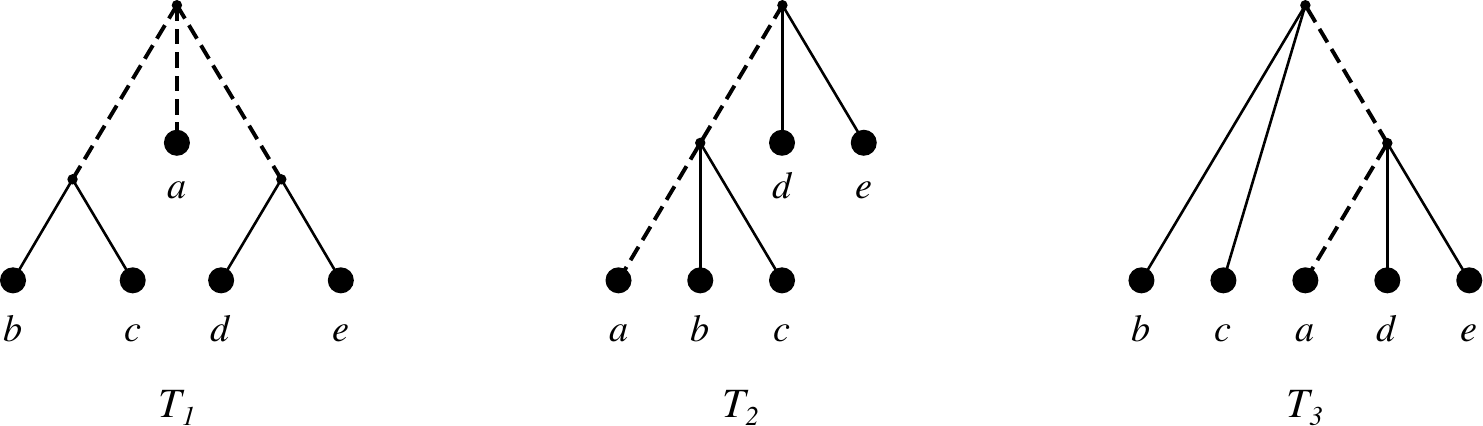}
  \end{center}
  \caption{Separating edges are indicated as dashed lines in the
    three trees with leaf set $X = \{a,b,c,d,e\}$, which explain the same
    partition $\mathcal{P} = \{\{a\}, \{b,c\}, \{d,e\}\}$.  $T_1$
    corresponds to the hierarchy
      $\mathcal{H}_{\mathcal{P}} \coloneqq \mathcal{P} \cup \{\{x\}\mid
      x\in X\} \cup \{X\}$.  The two trees $T_2$ and $T_3$ can be obtained
    from $T_1$ by contracting a single inner edge and re-rooting of the
    resulting tree.  In particular, $T_2$ and $T_3$ are minimally-resolved,
    that is, they have the fewest number of vertices among all trees that
    are compatible with $\mc{P}$. The last statement is easy to verify,
    since any tree with fewer vertices must be a star tree which is not
    compatible with $\mc{P}$.}
  \label{fig:minLRT}
\end{figure}

As we shall see in Lemma~\ref{lem:specialH}, we can always find a tree
  on $X$ that is compatible with $\mc{P}$ for a given partition $\mc{P}$ of
  $X$.  In particular, the tree corresponding to the hierarchy
  $\mathcal{H}_{\mathcal{P}} \coloneqq \mathcal{P} \cup \{\{x\}\mid x\in
  X\} \cup \{X\}$ is always compatible with $\mc{P}$, see $\mathcal{P}$ and
  $T_1$ in Fig.~\ref{fig:minLRT} for an illustrative example.
However, Fig.~\ref{fig:minLRT} also shows that there can be
multiple different trees on $X$ that are compatible with $\mc{P}$.
The main result of Section~\ref{sect:compatible}, Thm.~\ref{thm:compatible},
is a characterization of compatibility of partitions and hierarchies (and thus
rooted trees).

As it turns out, not all hierarchies are compatible with a given partition
$\mathcal{P}$.
In many applications, hierarchies (and their corresponding rooted trees) are
not
necessarily
fully resolved, even though it is often assumed that the ``ground truth'' is a
binary tree.
In situations where $\mathcal{H}$ and $\mathcal{P}$ are not compatible, it is
therefore of interest to ask whether it is possible to find a refinement of
$\mathcal{H}$ that is compatible with $\mathcal{P}$.
\begin{figure}[t]
  \begin{center}
    \includegraphics[width=.6\textwidth]{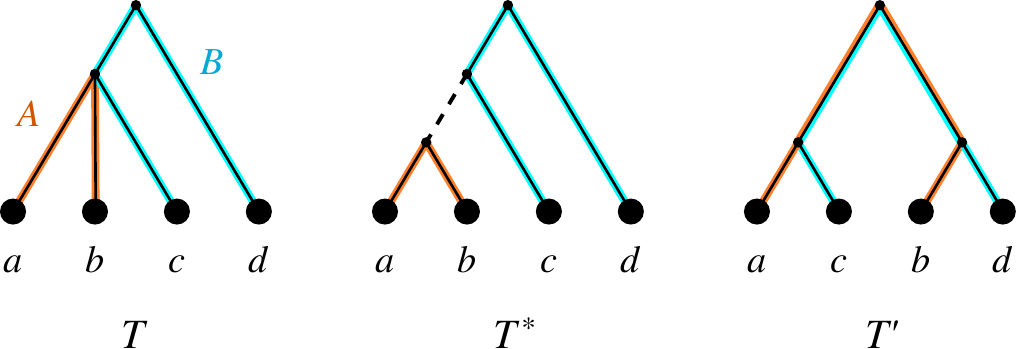}
  \end{center}
  \caption{The tree $T$ is not compatible with the partition
  $\mathcal{P}\coloneqq\{A\coloneqq\{a,b\}, B\coloneqq\{c,d\}\}$ but admits a
  compatible refinement $T^*$. In contrast, the
  tree $T'$ does not admit a compatible refinement. \emph{See text for more details.}}
  \label{fig:refinement-exists}
\end{figure}
An example of a tree $T$ that is not compatible with a partition
  $\mc{P}$ but that admits a compatible refinement $T^*$ is shown in
  Fig.~\ref{fig:refinement-exists}.  To see that $\mc{P}=\{A,B\}$ and $T$
  as in Fig.~\ref{fig:refinement-exists} are not compatible, the edges in
  $T$ have been colored in orange and cyan if they lie on a path connecting
  two elements from $A$ or $B$, respectively. Clearly, none of these edges
  can be a separating edge. In particular, since all edges in $T$ are
  colored, $A$ and $B$ cannot be separated by any subset $H\subseteq E(T)$.
  For similar reasons the tree $T'$ as in Fig.~\ref{fig:refinement-exists}
  is not compatible with $\mc{P}$. Since $T'$ is already fully-resolved, it
  does not admit a compatible refinement.
\begin{definition}\label{def:r-comp}
  A tree $T$ and a partition $\mathcal{P}$ are \emph{refinement-compatible}
  (\emph{r-compatible} for short) if there exists a refinement $T^*$ of $T$
  that is compatible with $\mathcal{P}$.

  In case $T$ is rooted (unrooted) and refinement-compatible with
  $\mathcal{P}$, the corresponding hierarchy $\mathcal{H}(T)$ (split system
  $\mathfrak{S}(T)$, resp.)  are said to be \emph{refinement-compatible}
  with $\mc{P}$.
\end{definition}
By definition, compatibility implies r-compatibility since every tree is a
refinement of itself.
In Section \ref{sect:refine}, we show that refining a hierarchy
$\mathcal{H}$ or a rooted tree $T$ that is already compatible with a
partition $\mathcal{P}$ never destroys compatibility.
As a main result of Section \ref{sect:refine}, we obtain a
characterization of r-compatibility in terms of the simple condition that
no set $Y\in \mathcal{H}$ overlaps with two distinct sets $A,B\in
\mathcal{P}$ (cf.\ Thm.~\ref{thm:compatible-refinement}).
We later utilize these results to derive simple linear-time algorithms for both
recognition of compatibility of a partition and a tree, as well as the
construction of a compatible refinement if
one exists in Section~\ref{sec:compl-algos}.

Even though compatibility of partitions and rooted trees is
linear-time-decidable, the situation appears substantially more complicated
when systems $\mathfrak{P}=\{\mathcal{P}_1,\mathcal{P}_2,\dots,\mathcal{P}_k\}$
of partitions of $X$ are considered rather than single partitions.
By definition, $T$ (or equivalently, the hierarchy,
$\mc{H}(T)$) and each $\mathcal{P}_i$ are compatible if and only if
$\mathcal{P}_i=\mathcal{F}(T,H_i)$ for some subset $H_i\subseteq E(T)$,
$1\leq i\leq k$. In this case, we say that $T$ (equiv.\ $\mc{H}(T)$) and
$\mathfrak{P}$ are compatible.
It is natural, then, to ask whether for a
given system of partitions $\mathfrak{P}$, there exists a
tree $T$ such that $\mathfrak{P}$ and $T$ are compatible:

  \begin{problem}[\textsc{Existence of Tree compatible with Partition System}
    (\textsc{ExistTP})]\ \\
    \begin{tabular}{ll}
      \emph{Input:}    & A partition system $\mathfrak{P}$ on $X$.\\
      \emph{Question:} & Is there a tree $T$ on $X$ such that
      $T$ and $\mathfrak{P}$ are compatible?\\
    \end{tabular}
  \end{problem}

Since every tree on $X$ is a refinement of the star tree on $X$,
\textsc{ExistTP} is a special case of the following more general problem:
\begin{problem}[\textsc{Compatibility of Tree and Partition System}
  (\textsc{CompaTP})]\ \\
  \begin{tabular}{ll}
    \emph{Input:}    & A tree $T$ with leaf set $X$ and a partition system
    $\mathfrak{P}$ on $X$.\\
    \emph{Question:} & Is there a refinement $T^*$ of $T$ such that
    $T^*$ and $\mathfrak{P}$ are compatible?\\
  \end{tabular}
\end{problem}

  \begin{figure}[t]
    \begin{center}
      \includegraphics[width=.8\textwidth]{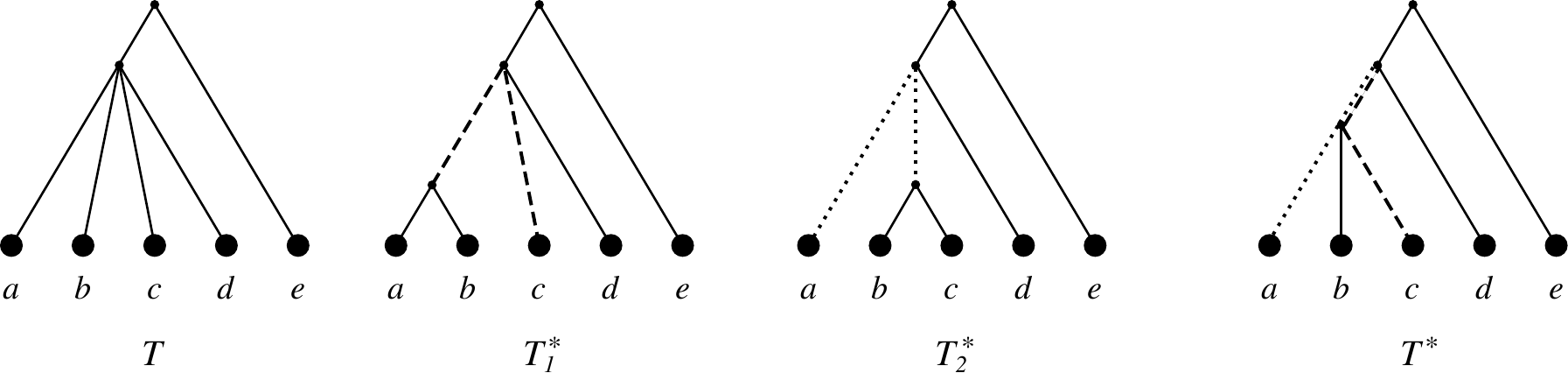}
    \end{center}
    \caption{The hierarchy $\mathcal{H}(T)$ corresponding to the tree $T$
      with leaf set $X=\{a,b,c,d,e\}$ that admits the refinements $T^*_1$ and
        $T^*_2$ of $T$ that are compatible with $\mathcal{P}_1 =
        \{\{a,b\},\{c\},\{d,e\}\}$ and $\mathcal{P}_2 =
        \{\{a\},\{b,c\},\{d,e\}\}$, respectively.
      The union $\mc{H}(T^*_1)\cup \mc{H}(T^*_2)$, however, does not form a
      hierarchy because the two elements $\{a,b\}$ and $\{b,c\}$ overlap.
      Thus, there is no common refinement of $T^*_1$ and $T^*_2$
      		that is compatible with $\mathfrak{P} =\{\mathcal{P}_1, \mathcal{P}_2\}$.
      In this example, $\mathcal{H}(T)$  has another refinement $\mc{H^*}$
      with corresponding tree $T^*$ that is compatible with both
      $\mathcal{P}_1$ and $\mathcal{P}_2$.}
    \label{fig:notHierarchy}
  \end{figure}
The difficulty of both \textsc{ExistTP} and \textsc{CompaTP} stems from
the fact that refinements of the underlying tree that are necessary to obtain
compatibility with individual partitions in $\mathfrak{P}$ may contradict one
another.
Consider  the tree $T$ and three of its possible refinements $T_1^*, T_2^*$ and
$T^*$ as shown in Fig.~\ref{fig:notHierarchy}. In this example, $T$ is not
compatible with $\mathfrak{P} = \{\mathcal{P}_1,\mathcal{P}_2\}$, and the
refinement $T_i^*$ is compatible with $\mathcal{P}_i$ but not with
$\mathcal{P}_j$, $\{i,j\}=\{1,2\}$.
However, there is no common refinement of  $T_1^*$ and $T_2^*$ that is
compatible with $\mathfrak{P}$.
On the other hand, $T^*$ is compatible with both $\mathcal{P}_1$ and
$\mathcal{P}_2$.
Hence, $T$ admits a refinement $T^*$ that is compatible with $\mathfrak{P}$.
An example of a tree  $T$ for which every partition $\mathcal{P}\in
\mathfrak{P}$ is r-compatible with $T$ but there is no refinement of $T$ that is
compatible with $\mathfrak{P}$ is provided in
Fig.~\ref{fig:no-common-refinement}.
Hence, \textsc{ExistTP} and \textsc{CompaTP} seem to be inherently difficult
and indeed both are NP-complete decision problems, see Thm.~\ref{thm:NPc}.

In Section~\ref{sec:splits}, compatibility of partitions with
splits systems and their equivalent representations as unrooted phylogenetic
trees is considered.
As shown in Prop.~\ref{prop:S-unroot}, compatibility of partitions with
unrooted and rooted trees are closely related. In fact, a partition is
compatible with an unrooted tree if and only if it is compatible with any
rooted version of this tree. Further characterization of (refinements of) split
systems and unrooted trees being compatible with partitions will be established
(cf. Thm.~\ref{thm:splitcomp} as well as Lemma \ref{lem:splitcompat} and
\ref{lem:splitrefine}).

Section~\ref{sec:compl-algos} is dedicated to algorithmic considerations
and the complexity of deciding whether (systems of) partitions are compatible
(with refinements) of hierarchies and split systems, represented by rooted and
unrooted trees, respectively.
As we have seen above, there are edges $e$ of a tree $T$ that can
never be separating edges since their removal would break
down some set $A\in\mc{P}$. In order to identify such edges, we provide an
edge coloring that we have already used in an informal way to demonstrate
incompatibility in the example in Fig.~\ref{fig:refinement-exists}:
\begin{definition}\label{def:P-col}
  Let $T$ be a tree on $X$ and $\mathcal{P}$ a partition of $X$.  The
  \emph{$\mathcal{P}$-(edge-)coloring of $T$} is the map
  $\gamma_{T,\mathcal{P}}\colon E(T) \to 2^{\mathcal{P}}$ that is given by
  \begin{equation*}
    A\in \gamma_{T,\mathcal{P}}(e) \iff e \text{ lies on the unique path
      connecting two } x,x'\in A.
  \end{equation*}
\end{definition}
The key property of this edge coloring $\gamma_{T,\mathcal{P}}$ is
that any edge $e\in E(T)$ with $\gamma_{T,\mathcal{P}}(e)\neq \emptyset$
lies on some path between two vertices $a,a'\in X$ that are contained in
the same set of $\mathcal{P}$, and thus, $e$ cannot be a separating edge.
In contrast, all edges $e$ for which $\gamma_{T,\mathcal{P}}(e)=\emptyset$
do not separate any two leaves that are in the same set of $\mathcal{P}$,
and thus can be safely added to the set of separating edges. A key result,
proven in Section~\ref{sec:compl-algos}, is that $\mc{P}$ and $T$ are
  r-compatible if and only if $|\gamma_{T,\mathcal{P}}(e)| \le 1$ for every
  $e\in E(T)$ (Prop.~\ref{prop:at-most-one-color}). In this case, the
  coloring $\gamma_{T,\mc{P}}$ can be computed in linear time and, based on
  this, deciding the existence of and finding a (refinement of) a tree that
  is compatible with a partition $\mathcal{P}$ can be in done in linear
  time as well. To establish compatibility of $\mc{P}$ and $T$, it
  suffices to rule out the existence of a vertex $u\in V(T)$ that is incident 
  with two differently colored edges (cf.\ 
  Thm.~\ref{thm:compatible-iff-color}). If $\mc{P}$ and $T$ are
  r-compatible but not compatible, the vertices $u\in V(T)$ that violate
  the latter condition coincide with $\lca_T(A)$ for some $A\in\mathcal{P}$
  and can be refined by collecting all children $v\in\child_{T}(u)$ for
  which $L(T(v))\cap A\ne\emptyset$ under a newly created vertex (cf.\
  Thm.~\ref{thm:comp-linear}).

\section{Compatibility of Partitions and Hierarchies}
\label{sect:compatible}

We first show that, for every partition, there is a compatible hierarchy.
\begin{lemma}
  For every partition $\mathcal{P}$ of $X$, the set system
  $\mathcal{H}_{\mathcal{P}} \coloneqq \mathcal{P} \cup \{\{x\}\mid x\in
  X\} \cup \{X\}$ is a hierarchy that is compatible with $\mathcal{P}$.
  In particular, every partition $\mathcal{P}$ of $X$ is compatible with a
  rooted tree on $X$.
  \label{lem:specialH}
\end{lemma}
\begin{proof}
  Since $\mathcal{P}$ is a partition of $X$, no two sets overlap.  This
  remains true for $\mathcal{H}_{\mathcal{P}}$ and thus
  $\mathcal{H}_{\mathcal{P}}$ is a hierarchy.  Moreover, it is easy to see
  that if all edges incident to the root of the tree $T$ corresponding to
  $\mathcal{H}$ are added to $H$, then $\mathcal{P} = \mathcal{F}(T,H)$
  whenever $\mathcal{P}\neq\{X\}$.  In case $\mathcal{P}=\{X\}$, $T$ is a
  star tree and we have $\mathcal{P} = \mathcal{F}(T,\emptyset)$.
\end{proof}

The following result is a key step for the characterization of compatible
partitions and hierarchies. In particular, it shows that, for any two
elements $A,B\in\mathcal{P}$, the set $B$ can only intersect with at most
one child of $\lca_T(A)$.
\begin{lemma}
  If a partition $\mathcal{P}$ of $X$ and a rooted tree $T$
  on $X$ are compatible,
  then, for all $A,B\in\mathcal{P}$, there are no two distinct children
  $u,u'\in \child(\lca_T(A))$ such that $B\cap L(T(u))\neq \emptyset$ and
  $B\cap L(T(u'))\neq \emptyset$.
  \label{lem:unique}
\end{lemma}
\begin{proof}
  Let $A\in\mathcal{P}$ and put $v_A \coloneqq \lca_T(A)$.  Since
  $\mathcal{P}$ is compatible with $\mathcal{H}$, there is a set of separating
  edges $H\subseteq E(T)$ such that $P=\mc{F}(T,H)$.  Assume, for
  contradiction, that there are two children $u,u'\in \child(v_A)$ and a
  set $B\in\mathcal{P}\setminus \{A\}$ such that
  $B\cap L(T(u))\neq \emptyset$ and $B\cap L(T(u'))\neq \emptyset$.  Since
  $v_A = \lca_T(A)$, there are children $w,w'\in\child(v_A)$ such that
  $A\cap L(T(w))\neq \emptyset$ and $A\cap L(T(w'))\ne \emptyset$.  The
  vertices $w,w'$ are not necessarily distinct from $u,u'$. However, we can
  assume w.l.o.g.\ that $u\ne w$. Since $\mathcal{P}$ is compatible with
  $\mathcal{H}$, there is no separating edge on the path from $a$ to $a'$ for
  any $a\in A\cap L(T(w))\ne\emptyset$ and $a' \in A\cap L(T(w'))\ne\emptyset$.
  In particular, the path from $a$ to $v_A$ does not contain a separating edge.
  Similarly, for all vertices $b\in B\cap L(T(u))\ne\emptyset$ and
  $b' \in B\cap L(T(u'))\ne\emptyset$, there is no separating edge on the path
  from $b$ to $b'$, and the path from any such $b$ to $v_A$ does not
  contain a separating edge. Since $u\ne w$ and both $u$ and $w$ are children of
  $v_A$, the path from $a$ to $b$ is the concatenation of the path from $a$
  to $v_A$ and the path from $b$ to $v_A$, and thus, does not contain a
  separating edge.  Hence $a$ and $b$ are in the same connected component of
  $\mathcal{F}(T,H)$; a contradiction since $A$ and $B$ are disjoint by
  assumption.
\end{proof}

\begin{figure}[t]
  \begin{center}
    \includegraphics[width=.3\textwidth]{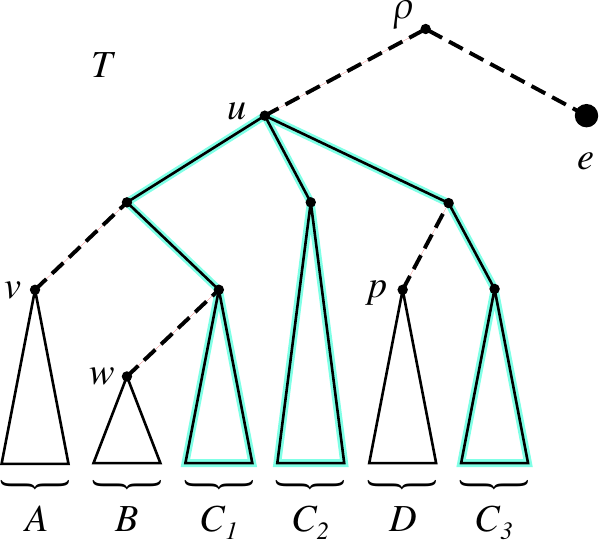}
  \end{center}
  \caption{The tree $T$ (on
    $X=\{A\cup B\cup C_1\cup C_2 \cup C_3 \cup D \cup \{e\}\}$) and the
    partition
    $\mathcal{P}=\{A,B,C\coloneqq C_1\cup C_2 \cup C_3, D,
    E\coloneqq\{e\}\}$ are compatible since removal of the separating edges
    (dashed-lined) that are chosen according to
    Eq.~\eqref{eq:hgt-edges}
    induces $\mathcal{P}$.  To emphasize its connectedness after removal of
    the separating edges, the minimal subtree connecting all elements in $C$ is
    highlighted in cyan.  Let $\mathcal{H}$ be the hierarchy corresponding
    to $T$.  For some $Y \in\mathcal{P}$, we have $Y=Y_{\mc{H}}$, that is,
    $L(T(v))=A_{\mathcal{H}}$ (i.e.\ the vertex $v$ corresponds to
    $A_{\mathcal{H}}\in\mathcal{H}$), $L(T(w))=B_{\mathcal{H}}$,
    $L(T(u))=C_{\mathcal{H}}$, and $L(T(e))=\{e\}=E_{\mathcal{H}}$.  For
    $C$ it holds that $C\subsetneq C_{\mc{H}} = L(T(u))$.  Moreover, some
    elements in $\mathcal{P}$ overlap with elements in $\mc{H}$, e.g.,
    $C\in \mathcal{P}$ overlaps with $D\cup C_3\in \mc{H}$.  Note that it
    suffices to have only one of $\{\rho_T,u\}$ and $\{\rho_T,e\}$ as a
    separating edge.}
  \label{fig:hier-part}
\end{figure}

Using the correspondence between phylogenetic trees and hierarchies (cf.\
Thm.~\ref{thm:1-1}), one
can translate Lemma~\ref{lem:unique} to the language of hierarchies:
\begin{corollary}
  Suppose a partition $\mathcal{P}$ of $X$ and a hierarchy $\mathcal{H}$
  on $X$ (with corresponding tree $T$) are compatible and $A,B\in\mathcal{P}$
  are distinct. Then $\lca_T(A)\neq \lca_T(B)$ and
  $A_{\mathcal{H}}=\cl(A)\neq \cl(B)=B_{\mathcal{H}}$.
\label{cor:unique}
\end{corollary}

\begin{lemma}
  Suppose a  partition $\mathcal{P}$ of $X$ and a hierarchy $\mathcal{H}$
  on $X$ are compatible, and $A\in\mathcal{P}$. The set $A_{\mathcal{H}}$ does
  not overlap with any $B\in \mathcal{P}$.
  \label{lem:noOverlap}
\end{lemma}
\begin{proof}
  Assume, for contradiction, that $A_{\mathcal{H}}$ and $B$ overlap for
  some $A,B\in \mathcal{P}$. Since $A\subseteq A_{\mathcal{H}}$ and
  $B\not\subseteq A_{\mathcal{H}}$, we have $A\neq B$ and, by
  Lemma~\ref{lem:unique}, $A_{\mathcal{H}}\neq B_{\mathcal{H}}$.  Since $B$
  overlaps with $A_{\mathcal{H}}$ and $\mathcal{H}$ is a hierarchy, we have
  $A_{\mathcal{H}}\subsetneq B_{\mathcal{H}}$. Now let $T$ be the tree of
  $\mathcal{H}$ and set
  $v_A\coloneqq \lca_T(A_\mc{H})\prec_T \lca_T(B_\mc{H}) \eqqcolon v_B$.
  Since $A_{\mathcal{H}}$ is the unique inclusion-minimal element in
  $\mathcal{H}$ that contains $A$, there are two distinct children
  $u,u'\in \child(v_A)$ such that $A\cap L(T(u))\neq \emptyset$ and
  $A\cap L(T(u'))\neq \emptyset$. Let $a\in A\cap L(T(u))$ and
  $a'\in A\cap L(T(u'))$. Since $\mathcal{P}$ and $\mathcal{H}$ are
  compatible, there is a set of separating edges $H\subseteq E(T)$ such that
  $\mathcal{P}=\mathcal{F}(T,H)$. Since $a,a'\in A$, the unique path
  $P_{a,a'}$ from $a$ to $a'$ in $T$, and in particular the path
  $P_{a,v_A}$ from $a$ to $v_A$, cannot contain any separating edge.
  Furthermore, since $A_{\mathcal{H}}$ and $B$ overlap, there is an element
  $b\in A_{\mathcal{H}}\cap B$. Since $A$ and $B$ are disjoint, we have
  $b\neq a,a'$. Hence, $b\preceq_T u''$ for some child
  $u''\in \child(v_A)$.  Since $u$ and $u'$ are distinct children of $v_A$,
  we can assume w.l.o.g.\ that $u''\ne u$.  Moreover, $v_A\prec_T v_B$
  together with the fact that $B_\mathcal{H}$ is the unique
  inclusion-minimal element in $\mathcal{H}$ that contains $B$, implies
  that there are two distinct children $w,w'\in \child(v_B)$ that satisfy
  $b\preceq_T u''\prec_T v_A\preceq_T w$ and $b'\in B\cap L(T(w'))$.  Since
  $b,b'\in B$, the unique path $P_{b,b'}$ from $b$ to $b'$ in $T$, and in
  particular the path $P_{b,v_A}$ from $b$ to $v_A$, cannot contain a
  separating edge.  Since $u\ne u''$ and both $u$ and $u''$ are children of
  $v_A$,  the path from $a$ to $b$ is the concatenation of the paths $P_{a,v_A}$
  and $P_{b,v_A}$, and thus, does not contain a separating edge.  Therefore,
  there is a set $C\in \mathcal{F}(T,H)$ with $a,b\in C$.  Now, $a\in A$
  and $b\in B$ implies that $\mathcal{P}\neq \mathcal{F}(T,H)$; a
  contradiction.
\end{proof}

\begin{theorem}
  Let $\mathcal{H}$ be a hierarchy on $X$ and $\mathcal{P}$ be a partition
  of $X$.  Then, $\mathcal{P}$ and $\mathcal{H}$ are compatible if and only
  if the following two conditions are satisfied for all
  $A,B\in\mathcal{P}$:
  \begin{enumerate}[noitemsep]
  \item[(i)] $A_\mathcal{H}$ is a union of sets of $\mathcal{P}$.
  \item[(ii)] If $A_\mathcal{H} = B_\mathcal{H}$, then $A=B$.
  \end{enumerate}
  \label{thm:compatible}
\end{theorem}
\begin{proof}
	Let $T$ be the tree with $\mathcal{H}(T)=\mathcal{H}$.

  Assume that $\mathcal{P}$ is compatible with $\mathcal{H}$.  First
  observe that, for all $A\in\mathcal{P}$, $A\subseteq
  A_{\mathcal{H}}$ implies
  that $A_{\mathcal{H}}$ is the union of sets of $\mathcal{P}$ if and only
  if $A_{\mathcal{H}}$ does not overlap with any $B\in\mathcal{P}$.  Hence,
  Condition~(i) follows immediately from Lemma~\ref{lem:noOverlap}.
  Condition (ii) follows immediately from the fact that
  $A_\mathcal{H} = B_\mathcal{H}$ implies $\lca_T(A)=\lca_T(B)$ and
  Cor.~\ref{cor:unique}.

  Now suppose (i) and (ii) holds. Consider the tree $T$ and the following
  set of separating edges
  \begin{equation}
    \label{eq:hgt-edges}
    H \coloneqq \left\{ \{\parent(\lca_T(A)),\lca_T(A)\} \mid A\in\mathcal{P},\
      \lca_T(A)\ne\rho_T \right\},
  \end{equation}
  Thus, an edge $e=\{u,v\}\in E(T)$ is a separating edge if and only if
  $L(T(v)) = A_{\mathcal{H}}$ and thus $v=\lca_T(A)$ for some
  $A\in\mathcal{P}$.

  We first show that any two distinct $A,B\in \mathcal{P}$ are separated by
  at least one separating edge in $H$, i.e., there is no path in $T-H$
  connecting
  any vertex $a\in A$ with any vertex $b\in B$.  To this end, let
  $A,B\in \mathcal{P}$ be chosen arbitrarily but distinct.  By
  contraposition of Condition~(ii), we have
  $A_{\mathcal{H}}\ne B_{\mathcal{H}}$.  Therefore and since $\mathcal{H}$
  is a hierarchy, we have to consider the two cases~(a)
  $A_{\mathcal{H}}\cap B_{\mathcal{H}}=\emptyset$, and~(b)
  $A_{\mathcal{H}}\subsetneq B_{\mathcal{H}}$ or
  $B_{\mathcal{H}}\subsetneq A_{\mathcal{H}}$.  Case~(a) corresponds to the
  situation in which $v_A\coloneqq \lca_T(A)$ and $v_B\coloneqq\lca_T(B)$
  are incomparable in $T$, which is, moreover, only possible if neither
  $v_A$ nor $v_B$ are the root.  Hence, the edges $\{\parent(v_A),v_A\}$
  and $\{\parent(v_B),v_B\}$ are contained in $H$ and every path from some
  $a\in A$ to some $b\in B$ contains these two edges. Thus, the two sets
  $A$ and $B$ are separated by separating edges in $T$.  In case~(b), we assume
  w.l.o.g.\ that $A_{\mathcal{H}}\subsetneq B_{\mathcal{H}}$ which
  corresponds to the situation in which
  $v_A\coloneqq\lca_T(A)\prec_T\lca_T(B)\eqqcolon v_B$.  Hence, $v_A$ is
  not the root of $T$, and we have $\{\parent(v_A),v_A\}\in H$.  Since
  $L(T(v_A))=A_{\mathcal{H}}$ contains all elements in $A$, the two sets
  $A$ and $B$ are completely separated by this separating edge if
  $A_{\mathcal{H}}$ does not contain any element of $B$.  Thus, assume for
  contradiction that $A_{\mathcal{H}}\cap B\ne \emptyset$.  This together
  with the facts that
  $A \subseteq A_{\mathcal{H}}\subsetneq B_{\mathcal{H}}$,
  $A\cap B= \emptyset$ and $B_{\mathcal{H}}$ is inclusion-minimal for $B$
  implies that $A_{\mathcal{H}}$ and $B$ overlap.
  Therefore, $A_{\mathcal{H}}$ is not
  the union of sets of $\mathcal{P}$; a contradiction to Condition~(i).

  It remains to show that no set $A\in P$ contains two elements $a,a'\in A$
  which are separated by a separating edge in $T$.  Thus, assume for
  contradiction that there is such an edge $e=\{u,v\}\in H$ lying on the
  path that connects two $a,a'\in A$ for some $A\in P$.  We can assume
  w.l.o.g.\ that $a\in L(T(v))$ and $a'\in L(T)\setminus L(T(v))$.  Since
  $e\in H$, we have that $v=\lca_T(B)$ corresponds to $B_{\mathcal{H}}$ for
  some $B\in\mathcal{P}$.  Since $a\in B_{\mathcal{H}}=L(T(v))$ but
  $a'\notin B_{\mathcal{H}}$ and by similar arguments as above, the sets
  $A$ and $B_{\mathcal{H}}$ overlap which is again a contradiction to
  Condition~(i).

  In summary, we have $A\in \mathcal{F}(T,H)=\mathcal{P}$ for all
  $A\in\mathcal{P}$ and thus, $\mathcal{P}=\mathcal{F}(T,H)$.
  Therefore, $\mathcal{P}$ is compatible with $\mathcal{H}$.
\end{proof}

\begin{figure}[t]
  \begin{center}
    \includegraphics[width=.75\textwidth]{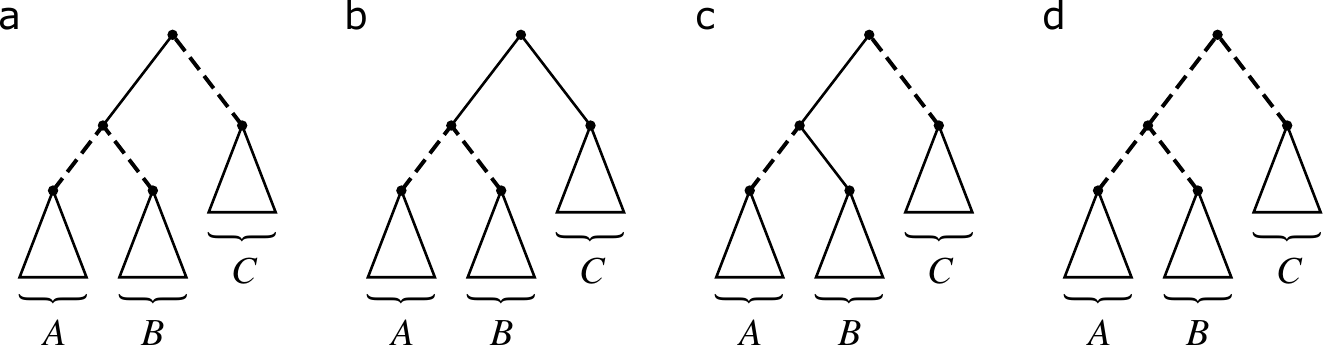}
  \end{center}
  \caption{The tree $T$ admits several possible choices for the set $H$ of
    separating edges (dashed-lined) such that
    $\mathcal{F}(T,H) = \mathcal{P}=\{A,B,C\}$. The separating edges in Panel
    (a) are the ones specified in Eq.~\eqref{eq:hgt-edges}. The two sets of
    separating edges in Panel (b) and (c) are both of minimum size, while the
    set of separating edges in Panel (d) is of maximum size. The set of
    separating edges of minimum size thus is not unique in general. In contrast,
    Cor.~\ref{cor:MaxHGT} implies that the maximum-sized set of separating
    edges
    is always unique. The set $H$ defined by Eq.~\eqref{eq:hgt-edges} is
    neither of minimum nor of maximum size.}
  \label{fig:minmaxHGT}
\end{figure}

The second part of the proof of Theorem~\ref{thm:compatible} implies a
simple algorithm to determine whether $\mathcal{P}$ and $\mathcal{H}$ are
compatible and to construct a (minimal) set $H$ of separating edges
that realizes the partition $\mathcal{P}$ on $T$:
In Section~\ref{sec:compl-algos}, we derive a linear-time compatibility test
algorithm.
The set $H$ of separating edges in Eq.~\eqref{eq:hgt-edges} can also be
constructed in polynomial time.
Moreover, if there is an element $A\in\mathcal{P}$ with
$A_{\mathcal{H}}=X$, then the set $H$ as in Eq.~\eqref{eq:hgt-edges} is
minimal because it contains, by construction, $|\mathcal{P}|-1$ separating
edges, i.e., the minimal number of splits required to decompose a tree into
$|\mathcal{P}|$ connected components. If there is no $A\in\mathcal{P}$ with
$A_{\mathcal{H}}=X$, however, then there is one edge too many. A minimal
set of separating edges can easily obtained in this case by omitting the
separating edge $\{\parent(\lca_T(A)),\lca_T(A)\}$ for one of the sets that are
inclusion-maximal among the inclusion-minimal sets $A_{\mathcal{H}}$, i.e.,
those which correspond to vertices that are closest to the root (see Fig.\
\ref{fig:hier-part} and \ref{fig:minmaxHGT} for further examples).
We summarize the latter discussion in the following
\begin{lemma}
  Suppose that a partition $\mathcal{P}$ of $X$ and a rooted
  tree $T$ on $X$ are compatible.
 Then, there always
  exists a minimum-sized set of separating edges $H^*$ such that
  $\mathcal{P} = \mathcal{F}(T,H^*)$ and $|H^*| = |\mathcal{P}|-1$.  In
  particular, if $H$ is chosen as in Eq.~\eqref{eq:hgt-edges}, then
  $|H|\in \{|\mathcal{P}|-1, |\mathcal{P}|\}$ and $|H| = |\mc P|-1$ if and
  only if there is no $A\in\mathcal{P}$ with $A_{\mathcal{H}}=X$.
  \label{lem:min-size}
\end{lemma}

The following result is a simple consequence of Thm.~\ref{thm:compatible}
and the discussion above.
\begin{corollary}
  Let $\mathcal{P}$ be a partition of $X$, $\mathcal{H}$ a hierarchy
  on $X$ with corresponding tree $T$, and let $H$ be the edge set
  defined in Eq.~\eqref{eq:hgt-edges}. Then $\mathcal{P}$ and
  $\mathcal{H}$ are compatible if and only if
  $\mathcal{P} = \mathcal{F}(T,H)$.
  \label{cor:set-H}
\end{corollary}

\begin{lemma}
  If the partition $\mathcal{P}$ and the hierarchy $\mathcal{H}$ on $X$ are
  compatible, then the following conditions hold for all
  $A,B\in\mathcal{P}$: If $B\subseteq A_\mathcal{H}$ and $B\ne A$, then
  $B_\mathcal{H}\cap A=\emptyset$.
  \label{lem:claim3}
\end{lemma}
\begin{proof}
  By Property (i) of Thm.~\ref{thm:compatible}, $B_{\mathcal{H}}$ is a
  union of sets of $\mathcal{P}$, and thus either (a)
  $A\subseteq B_{\mathcal{H}}$ or (b) $A\cap B_{\mathcal{H}}=\emptyset$. In
  case (a), we have
  $A_{\mathcal{H}}=\cl(A)\subseteq \cl(B_{\mathcal{H}})=B_{\mathcal{H}}$ by
  isotony and idempotence of the closure.  Similarly, we obtain
  $B_{\mathcal{H}}\subseteq A_{\mathcal{H}}$ from the assumption
  $B\subseteq A_{\mathcal{H}}$. Therefore,
  $A_{\mathcal{H}}=B_{\mathcal{H}}$. By Property (ii) of
  Thm.~\ref{thm:compatible} this is a contradiction to $A\ne B$.  Thus,
  case~(a) is impossible and we always have
  $A\cap B_{\mathcal{H}}=\emptyset$.
\end{proof}
Thm.~\ref{thm:compatible} and Lemma~\ref{lem:claim3} together can be
rephrased as
\begin{corollary}
  The partition $\mathcal{P}$ and the hierarchy $\mathcal{H}$ are
  compatible if and only if
  \begin{equation}
    A = A_{\mathcal{H}} \setminus \bigcup_{\substack{B\in\mathcal{P}\\
        B_{\mathcal{H}}\subsetneq A_{\mathcal{H}}}} B_{\mathcal{H}}
  \end{equation}
  holds for all $A\in\mathcal{P}$.
\end{corollary}

\section{Compatibility of (Systems of) Partitions and Refinements of
Hierarchies}
\label{sect:refine}

In many applications hierarchies (and their associated trees) are not
necessarily fully resolved, even though it is often assumed that the
``ground truth'' is a binary tree.
 We show first that refining a hierarchy
$\mathcal{H}$ or a tree $T$ that is already compatible with a partition
$\mathcal{P}$ never destroys compatibility.
\begin{proposition}
  \label{prop:refinement}
  A hierarchy $\mathcal{H}$ on $X$ and a
  partition $\mathcal{P}$ of $X$ are compatible if and only if
  $\mathcal{P}$ is compatible with every refinement $\mathcal{H}^*$ of
  $\mathcal{H}$.
\end{proposition}
\begin{proof}
	The \emph{if}-direction immediately follows from the fact that
	$\mathcal{H}^*=\mathcal{H}$
	is a refinement of $\mathcal{H}$. For the \emph{only-if}-direction,
  let $T$ be the tree corresponding to $\mathcal{H}$, denote by
  $H\subseteq E(T)$ the set of separating edges as defined in
  Eq.~\eqref{eq:hgt-edges}.  Thus, we have
  $e=\{\parent_T(v),v\}\in E(T)\cap H$ if and only if
  $v=\lca_T(A)\ne \rho_T$.  Since $\mathcal{H}$ and $\mathcal{P}$ are
  compatible, we can apply Cor.~\ref{cor:set-H} to conclude that
  $\mathcal{P}=\mathcal{F}(T,H)$.  Therefore, the path connecting any two
  vertices $a\in A$ and $b\in B$ from distinct $A,B\in\mathcal{P}$ contains
  at least one edge in $H$.  Let $\mathcal{Y}$ be the set of all
  $Y\in\mathcal{H}$ with $v = \lca_T(Y)$ and $\{\parent_T(v),v\}\in H$. For
  all $Y\in\mathcal{Y}\subseteq\mathcal{H}\setminus\{X\}$, therefore, there
  is an $A\in\mathcal{P}$ such that $Y=A_{\mathcal{H}}$.

  Now consider an arbitrary refinement $\mathcal{H}^*$ of $\mathcal{H}$ and
  the corresponding refinement $T^*$ of $T$. By construction, we have
  $\mathcal{Y}\subset\mathcal{H}\subseteq\mathcal{H}^*$. For each
  $Y\in\mathcal{Y}$, we set $v_y\coloneqq \lca_{T^*}(Y)$ and set
  \begin{equation*}
    H^* \coloneqq \left\{ \{\parent_{T^*}(v_Y),v_Y\} \mid
      Y\in\mathcal{Y}\right\}.
  \end{equation*}
  Since $X\notin \mathcal{Y}$ by construction, we have $v_Y\ne\rho_{T^*}$
  and thus $\parent_{T^*}(v_Y)$ and, in particular, $H^*$ is well-defined.

  Now consider two arbitrary two vertices $a\in A$ and $b\in B$ in distinct
  $A,B\in\mathcal{P}$. As argued above, the path connecting them in $T$
  contains an edge $\{\parent_{T}(v),v\}\in H$.  By construction, we have
  $Y\coloneqq L(T(v))\in \mathcal{Y}$.  We can assume w.l.o.g.\ that
  $a\in Y$ and $b\in X\setminus Y$.  This together with $Y\in\mathcal{H}^*$
  implies that the path connecting $a$ and $b$ in $T^*$ contains the edge
  $\{\parent_{T^*}(v_Y),v_Y\}$.  By construction of $H^*$ and since
  $Y\in\mathcal{Y}$, we have $\{\parent_{T^*}(v_Y),v_Y\}\in H^*$.

  It remains to show that the path in $T^*$ connecting any two $a,a'\in A$
  for some $A\in\mathcal{P}$ never contains an edge that is in $H^*$.
  Assume, for contradiction, that this is the case, i.e., there is an edge
  $\{\parent_{T^*}(v),v\}\in H^*$ such that w.l.o.g.\
  $a\in L(T(v))\eqqcolon Y'$ and $a'\in X\setminus Y'$.  By construction of
  $H^*$, $Y'\in\mathcal{Y}$ and $Y'=B_{\mathcal{H}}$ for some
  $B\in\mathcal{P}$.  Since $\mathcal{H}$ and $\mathcal{P}$ are compatible,
  Thm.~\ref{thm:compatible}(i) implies that $Y'=B_{\mathcal{H}}$ is the
  union of sets of $\mathcal{P}$; a contradiction to the fact that
  $a\in Y'$ and $a'\notin Y'$.

  In summary, we conclude that $\mathcal{P}=\mathcal{F}(T^*,H^*)$, and,
  therefore, $\mathcal{P}$ is compatible with $\mathcal{H}^*$.
\end{proof}

  We next provide  a necessary condition for
  r-compatibility.  We
  will show later that this condition is also sufficient.
\begin{lemma}
  \label{lem:refinement-implies-no-overlaps}
  Let $\mathcal{H}$ be a hierarchy on $X$ and
  $\mathcal{P}$ a partition of $X$.  If $\mathcal{P}$ is
  compatible with a refinement $\mathcal{H}^*$ of $\mathcal{H}$, then there is
  no set $Y\in \mathcal{H}$ that overlaps with two distinct sets $A,B\in
  \mathcal{P}$.
\end{lemma}
\begin{proof}
  Let $\mathcal{P}$ and $\mathcal{H}^*$ with
  $\mathcal{H}\subseteq \mathcal{H}^*$ be compatible.  Assume, for
  contradiction, that some $Y\in \mathcal{H}$ overlaps with two distinct
  $A,B\in \mathcal{P}$.  First observe that $Y\in\mathcal{H}^*$.  Now
  consider $A_{\mathcal{H}^*}$ and $B_{\mathcal{H}^*}$.  Since $A\ne B$ and
  $\mathcal{P}$ is compatible with $\mathcal{H}^*$, we have
  $A_{\mathcal{H}^*}\ne B_{\mathcal{H}^*}$ by contraposition of
  Thm.~\ref{thm:compatible}(ii).  Since $Y$ overlaps with $A$ and $B$, both
  $A$ and $B$, and thus $A_{\mathcal{H}^*}$ and $B_{\mathcal{H}^*}$,
  contain elements that are in $Y$ and as well as elements that are not in
  $Y$. This together with the fact that $Y$, $A_{\mathcal{H}^*}$, and
  $B_{\mathcal{H}^*}$ are all sets in $\mathcal{H}^*$ implies that
  $Y\subsetneq A_{\mathcal{H}^*}$ and $Y\subsetneq B_{\mathcal{H}^*}$.
  Therefore we have $Y\subseteq A_{\mathcal{H}^*} \cap B_{\mathcal{H}^*}$.
  Since $Y\ne\emptyset$, $\mathcal{H}^*$ is a hierarchy, and
  $A_{\mathcal{H}^*}\ne B_{\mathcal{H}^*}$, this implies that either
  $A_{\mathcal{H}^*}\subsetneq B_{\mathcal{H}^*}$ or
  $B_{\mathcal{H}^*}\subsetneq A_{\mathcal{H}^*}$.  Assume w.l.o.g.\ that
  $A_{\mathcal{H}^*}\subsetneq B_{\mathcal{H}^*}$.  Since $Y$ overlaps with
  $B$ and $Y\subsetneq A_{\mathcal{H}^*}$, we have
  $A_{\mathcal{H}^*}\cap B \ne \emptyset$.  However, since
  $A_{\mathcal{H}^*}\subsetneq B_{\mathcal{H}^*}$ and $B_{\mathcal{H}^*}$
  is inclusion-minimal for $B$ in the hierarchy $\mathcal{H}^*$, we
  conclude that $B$ and $A_{\mathcal{H}^*}$ overlap.  Hence,
  $A_{\mathcal{H}^*}$ is not the union of sets of $\mathcal{P}$. This
  violates Condition~(i) for compatible partitions in
  Thm.~\ref{thm:compatible}; a contradiction.
\end{proof}
To show that the converse of
  Lemma~\ref{lem:refinement-implies-no-overlaps} is satisfied as well, we
  will explicitly construct a compatible refinement
  of $\mathcal{H}$. To this end, we introduce the
following subset of a partition $\mathcal{P}$:
\begin{equation}
  \mathfrak{Y}(\mathcal{H},\mathcal{P})\coloneqq \{ A\in\mathcal{P} \mid
  \exists B\in\mathcal{P}\setminus\{A\} \text{ with } B\cap
  A_{\mathcal{H}}\ne\emptyset \text{ and } A_{\mathcal{H}}\subseteq
  B_{\mathcal{H}}\}
  \label{eq:unresolved-sets}
\end{equation}
The set $\mathfrak{Y}(\mathcal{H},\mathcal{P})$ contains the sets
$A\in\mathcal{P}$ for which $A_{\mathcal{H}}\in \mathcal{H}$ or, equivalently,
the vertex $u = \lca_T(A)$ is ``not resolved enough'':
\begin{proposition}
  A hierarchy $\mathcal{H}$ on $X$ and a partition $\mathcal{P}$
  of $X$ are compatible if and only
  if $\mathfrak{Y}(\mathcal{H},\mathcal{P})$ is empty.
  \label{prop:compatible-single-cond}
\end{proposition}
\begin{proof}
  By Thm.~\ref{thm:compatible}, $\mathcal{H}$ and $\mathcal{P}$ are compatible
  if and only if, for all $A,B\in\mathcal{P}$,
  the following two conditions are satisfied: (i) $A_\mathcal{H}$ is a union of
  sets of $\mathcal{P}$, and (ii) $A_\mathcal{H} = B_\mathcal{H}$ implies $A=B$.
  Hence, it suffices to show that
  $\mathfrak{Y}(\mathcal{H},\mathcal{P})=\emptyset$ is equivalent to these two
  conditions.

  First assume, for contraposition, that
  $\mathfrak{Y}(\mathcal{H},\mathcal{P})$ is not empty.
  Hence, there are distinct $A,B\in\mathcal{P}$ such that $B\cap
  A_{\mathcal{H}}\ne\emptyset$ and $A_{\mathcal{H}}\subseteq
  B_{\mathcal{H}}$.
  If $A_{\mathcal{H}}=B_{\mathcal{H}}$, then Condition~(ii) is violated.
  If on the other hand $A_{\mathcal{H}}\subsetneq B_{\mathcal{H}}$, then the
  fact that $B_{\mathcal{H}}$ is inclusion-minimal for $B$ implies that
  $B\setminus A_{\mathcal{H}}\ne\emptyset$.
  This together with $B\cap A_{\mathcal{H}}\ne\emptyset$ in turn implies that
  $A_\mathcal{H}$ is not a union of sets of $\mathcal{P}$; a violation of
  Condition~(i).

  The converse can be shown by very similar arguments starting with the
  assumption that Condition~(i) or~(ii) is not satisfied. If Condition~(i)
  does not hold, then $A_{\mc{H}}\neq X$ and, in particular, there is a
  $B\in \mc P$ such that $A_{\mc{H}}$ and $B$ overlap. Since $H$ is a
  hierarchy, it holds that $A_{\mc{H}}\subsetneq B_{\mc{H}}\subseteq X$.
  The latter two arguments imply
  $\mathfrak{Y}(\mathcal{H},\mathcal{P})\neq \emptyset$ since it contains
  $A$.  If Condition (ii) does not hold, then there are two distinct
  elements $A,B\in \mc{P}$ with $A_{\mc H}=B_{\mc H}$. Hence,
  $B\subseteq B_{\mc H}$ implies $B\cap A_{\mc H}\neq \emptyset$.  Thus,
  $A\in \mathfrak{Y}(\mathcal{H},\mathcal{P})$.
\end{proof}
In particular, the set $\mathfrak{Y}(\mathcal{H},\mathcal{P})$ can be used
to characterize the cases in which a compatible refinement of $\mathcal{H}$
exists and, if this is the case, to construct such a refinement.
\begin{definition}
  Let $\mathcal{H}$ be a hierarchy on $X$ and
  $\mathcal{P}$ a partition of $X$ such that no set $Y\in \mathcal{H}$ overlaps
  with two distinct sets $A,B\in \mathcal{P}$.
  We define, for $A\in\mathfrak{Y}(\mathcal{H},\mathcal{P})$, the subset $Y_A$
  of $A_{\mathcal{H}}$ as
  \begin{equation*}
    Y_A\coloneqq W_1 \cupdot W_2 \cupdot\dots\cupdot W_k
  \end{equation*}
  where $W_1,W_2,\dots W_k\in\mathcal{H}$ are the child clusters of
  $A_{\mathcal{H}}$ for which $W_i\cap A\ne\emptyset$.
  Moreover, the subset $\HP$ of $2^X$ is given by
  \begin{equation*}
    \HP \coloneqq \mathcal{H} \cup \{Y_A \mid
    A\in\mathfrak{Y}(\mathcal{H},\mathcal{P})\}.
  \end{equation*}
\end{definition}
Note that,
by Eq.~\eqref{eq:unresolved-sets}, for every
$A\in\mathfrak{Y}(\mathcal{H},\mathcal{P})$, the set
$A_{\mathcal{H}}$ cannot be a singleton.
Since $A_{\mathcal{H}}$ is inclusion-minimal w.r.t.\ $A$ by definition, we
immediately conclude that $k\ge 2$ and thus the $W_i$ are proper subsets of
$Y_A$.  In terms of the tree $T$ corresponding to $\mathcal{H}$, the
vertices $w_i\coloneqq \lca_T(W_i)$ are the children of
$y\coloneqq \lca_T(A_{\mathcal{H}})$ with $A\cap L(T(w_i))\ne\emptyset$.
\begin{lemma}
  \label{lem:HP-properties}
  Let $\mathcal{H}$ be a hierarchy on $X$ and $\mathcal{P}$ a partition of $X$
  such that no set $Y\in \mathcal{H}$ overlaps with two distinct sets $A,B\in
  \mathcal{P}$.
  Then the following two statements are satisfied:
  \begin{enumerate}
    \item For each $A\in\mathfrak{Y}(\mathcal{H},\mathcal{P})$, it holds
    $Y_A\notin \mathcal{H}$ and, in
    particular, $Y_A\subsetneq A_{\mathcal{H}}$.
    \item The set $\HP$ is a hierarchy.
  \end{enumerate}
\end{lemma}
\begin{proof}
  To show (1), set
    $A\in\mathfrak{Y}\coloneqq\mathfrak{Y}(\mathcal{H},\mathcal{P})$.
    By construction, we have $Y_A\subseteq A_{\mathcal{H}}$.
  Assume for contradiction that
  $Y_A=A_{\mathcal{H}}$, i.e., all child clusters
  of
  $A_{\mathcal{H}}$ in $\mathcal{H}$ have a non-empty intersection with
  $A$.  By definition of $\mathfrak{Y}$, we must have $|A_{\mathcal{H}}|>1$
  and, in particular, $A_{\mathcal{H}}$ has a
   child cluster $Y'\subsetneq A_{\mathcal{H}}$ satisfying
  $B\cap Y'\ne\emptyset$ for some $B\in\mathcal{P}\setminus\{A\}$ such that
  $A_{\mathcal{H}}\subseteq B_{\mathcal{H}}$. Thus
  $Y'\setminus A\ne \emptyset$ and $A\setminus Y'\ne\emptyset$, i.e., $A$
  and $Y'$ overlap.  Since
  $Y'\subsetneq A_{\mathcal{H}}\subseteq B_{\mathcal{H}}$ and
  $B_{\mathcal{H}}$ is inclusion-minimal w.r.t.\ $B$, we conclude that
  $B\setminus Y'\ne\emptyset$. On the other hand, since
  $Y'\cap A\ne\emptyset$ and $A,B\in\mathcal{P}$ are disjoint, we have
  $Y'\setminus B\ne\emptyset$. Thus $B$ and $Y'$ overlap. Thus there are
  two distinct sets $A,B\in\mathcal{P}$ that overlap with
  $Y'\in\mathcal{H}$. This contradicts the assumption that no such pair of
  sets exists, hence $Y_A\ne A_{\mathcal{H}}$.  Since $Y_A\subseteq Y$ by
  construction, $Y_A$ is a proper subset of $A_{\mathcal{H}}$.
  This together with the fact that $\mathcal{H}$ is a hierarchy and $Y_A$
  is the union of at least two child clusters
  of $A_{\mathcal{H}}$ in $\mathcal{H}$
  implies that $Y_A\notin \mathcal{H}$.

  We proceed by showing that the set system
  $\HP=\mathcal{H}\cup\{Y_A \mid A\in\mathfrak{Y}\}$ is again a
  hierarchy and thus, that (2) is satisfied.  To this end, consider first
  one of the newly-created sets
  $Y_A= W_1 \cupdot W_2 \cupdot\dots\cupdot W_k$ and an arbitrary set
  $Y'\in\mathcal{H}$.  If $W_i\cap Y'=\emptyset$ for all $1\le i\le k$,
  then $Y_A\cap Y'=\emptyset$.  Otherwise, there is some $W_i$ such that
  $W_i\cap Y'\ne\emptyset$.  Since both $W_i$ and $Y'$ are sets of the
  hierarchy $\mathcal{H}$, this implies that
  $Y'\subseteq W_i\subsetneq Y_A$, or $W_i\subsetneq Y'$.  In the latter
  case, we have $A_{\mathcal{H}}\subseteq Y'$ by the hierarchy property of
  $\mathcal{H}$ and the fact that $W_i$ is a child cluster of
  $A_{\mathcal{H}} (\in \mathcal{H})$. Hence, we have
  $Y_A\cap Y'\in\{\emptyset,Y_A,Y'\}$ for all $Y'\in\mathcal{H}$.  Now
  consider two newly-created sets $Y_A$ and $Y_B$ for distinct
  $A,B\in\mathfrak{Y}$, and assume first that
  $A_{\mathcal{H}}\ne B_{\mathcal{H}}$.  If
  $A_{\mathcal{H}}\cap B_{\mathcal{H}}=\emptyset$, then
  $Y_A\subsetneq A_{\mathcal{H}}$ and $Y_B\subsetneq B_{\mathcal{H}}$
  immediately imply that $Y_A\cap Y_B=\emptyset$.  Otherwise, we can assume
  w.l.o.g.\ that $B_{\mathcal{H}}\subsetneq A_{\mathcal{H}}$ since
  $A_{\mathcal{H}}$ and $B_{\mathcal{H}}$ are both sets in the hierarchy
  $\mathcal{H}$.  This together with the fact that $Y_A$ is the union of
  child clusters $W_1,W_2,\dots,W_k\in\mathcal{H}$
  of $A_{\mathcal{H}}$  implies that either
  $B_{\mathcal{H}}\subseteq W_i$ for some $1\le i\le k$, and thus
  $Y_B \subsetneq B_{\mathcal{H}} \subseteq W_i \subsetneq Y_A$, or, if no
  such $W_i$ exists, $Y_B\cap Y_A=\emptyset$.  It remains to consider two
  distinct sets $A,B\in\mathfrak{Y}$ with
  $A_{\mathcal{H}}=B_{\mathcal{H}}$.  If $Y_A$ and $Y_B$ overlap, then
  there is by construction a child cluster
  $W'$ of $A_{\mathcal{H}}=B_{\mathcal{H}}$ in $\mathcal{H}$ such that
  $W'\cap A\ne\emptyset$ and $W'\cap B\ne\emptyset$. Since, moreover,
  $A_{\mathcal{H}}=B_{\mathcal{H}}$ is inclusion-minimal w.r.t.\ $A$ and
  $B$, this implies that $W'\in\mathcal{H}$ overlaps with both $A$ and $B$;
  a contradiction to the assumption. Thus $Y_A$ and $Y_B$ are disjoint.  In
  summary, no two sets in $\HP$ overlap. Since
  $\mathcal{H}\subseteq\HP$, $X\in\mathcal{H}$ and
  $\{x\}\in\mathcal{H}$ for all $x\in X$, we conclude that $\HP$
  is a hierarchy that refines $\mathcal{H}$.
\end{proof}

The final step towards characterizing r-compatibility is a sufficient
  condition for $\HP$ to be compatible with $\mathcal{P}$.
\begin{lemma}
  \label{lem:HP-compatible}
  Let $\mathcal{H}$ be a hierarchy on $X$ and $\mathcal{P}$ a
    partition of $X$. If there is no set $Y\in \mathcal{H}$ that overlaps
    with two distinct sets $A,B\in\mathcal{P}$, then the hierarchy $\HP$ is
    compatible with $\mathcal{P}$.  
\end{lemma}
\begin{proof}
  Recall that $\HP$ is a hierarchy by Lemma~\ref{lem:HP-properties}(2).
	  To prove that $\HP$ is compatible with $\mathcal{P}$, we show that
		the Condition (i) and (ii) in Thm.~\ref{thm:compatible} are satisfied
		for $\HP$ and $\mathcal{P}$.

  To show Condition~(i) in Thm.~\ref{thm:compatible}, we assume, for
  contradiction, that there is a set $A\in\mathcal{P}$ for which
  $A_{\HP}$ is not the union of sets of $\mathcal{P}$. Thus
  there is a set $B\in\mathcal{P}\setminus \{A\}$ such that
  $A_{\HP}\cap B\ne\emptyset$ and
  $B\setminus A_{\HP}\ne \emptyset$.  We distinguish the two
  cases~(1) $A_{\HP}\notin \mathcal{H}$ and
  (2)~$A_{\HP}\in \mathcal{H}$.\\
  In case~(1), we have $A_{\HP}=Y_{A'}$ for some
  $A'\in\mathfrak{Y}$.  Thus $A_{\HP}$ is the union
  $W_1 \cupdot W_2 \cupdot\dots\cupdot W_k$ of $k\ge 2$ sets
  $W_1,W_2,\dots,W_k\in \mathcal{H}$ all satisfying
  $W_i\cap A'\ne\emptyset$.  Since
  $B\setminus A_{\HP}\ne \emptyset$ and by construction
  $A'\subseteq Y_{A'}=A_{\HP}$, we have $A'\ne B$.  Since
  $A_{\HP}\cap B\ne\emptyset$, there must be some
  $W'\in\{W_1,W_2,\dots,W_k\}\subset \mathcal{H}$ such that
  $B\cap W'\ne\emptyset$.  Since $W'\subsetneq A_{\HP}$ and
  $B\setminus A_{\HP}\ne \emptyset$, we have
  $B\setminus W'\ne\emptyset$.  On the other hand, we also have
  $A'\setminus W'\ne\emptyset$ because $k\ge 2$ and
  $W_i\cap A'\ne\emptyset$ for all $1\le i\le k$. Since $A'$ and $B$ are
  disjoint and both have a
  non-empty intersection with $W'$, we also obtain $W'\setminus
  A'\ne\emptyset$ and
  $W'\setminus B\ne\emptyset$.  In summary, the set $W'\in \mathcal{H}$
  overlaps with the two distinct sets $A',B\in\mathcal{P}$; a contradiction
  to the assumption.\\
  In case~(2), we have $A_{\HP}\in \mathcal{H}$.  Since
  $A_{\HP}$ is inclusion-minimal for $A$ in $\HP$ and
  $\mathcal{H}\subseteq\HP$ we conclude that $A_{\HP}$
  is also inclusion-minimal for $A$ in $\mathcal{H}$, and thus
  $A_{\HP}=A_{\mathcal{H}}$.  Since
  $A_{\HP}\cap B\ne\emptyset$,
  $B\setminus A_{\HP}\ne \emptyset$, and $\mathcal{H}$ is a
  hierarchy, we conclude that $A_{\mathcal{H}}\subsetneq B_{\mathcal{H}}$.
  In summary, we obtain $A\in \mathfrak{Y}$.  Hence, we
  have added a set $Y_A$ that satisfies $A\subseteq Y_A$ and, by the
  arguments above, $Y_A\subsetneq A_{\HP}$.  Therefore,
  $A_{\HP}$ is not inclusion-minimal for $A$ in $\HP$;
  a contradiction.

  To show Condition~(ii) in Thm.~\ref{thm:compatible}, we assume, for
  contradiction, that $Y^*\coloneqq A_{\HP} = B_{\HP}$
  for two distinct $A,B\in\mathcal{P}$.  As above, we distinguish the two
  cases~(1') $Y^*\notin \mathcal{H}$ and
  (2')~$Y^*\in \mathcal{H}$.\\
  In case~(1'), we have $Y^*=Y_{A'}$ for some $A'\in\mathfrak{Y}$.
  Thus $Y^*$ is the union $W_1 \cupdot W_2 \cupdot\dots\cupdot W_k$ of
  $k\ge 2$ sets $W_1,W_2,\dots,W_k\in \mathcal{H}$ all satisfying
  $W_i\cap A'\ne\emptyset$.  Since $A$ and $B$ are distinct, we can assume
  w.l.o.g.\ that $A'\ne B$.  Since $Y^*\cap B\ne\emptyset$, there must be
  some $W'\in\{W_1,W_2,\dots,W_k\}\subset \mathcal{H}$ such that
  $B\cap W'\ne\emptyset$.  Since $W'\subsetneq Y^*$ and $Y^*$ is
  inclusion-minimal for $B$ in $\HP$, we have
  $B\setminus W'\ne\emptyset$.  On the other hand, we also have
  $A'\setminus W'\ne\emptyset$ since $k\ge 2$ and $W_i\cap
  A'\ne\emptyset$ for all $1\le i\le k$. Since $A'$ and $B$ are disjoint
  and both have a
  non-empty intersection with $W'$, we also obtain $W'\setminus
  A'\ne\emptyset$ and
  $W'\setminus B\ne\emptyset$.  In summary, the set $W'\in \mathcal{H}$
  overlaps with the two distinct sets $A',B\in\mathcal{P}$; a contradiction
  to the assumption.\\
  In case~(2'), we have $Y^*\in \mathcal{H}$. Together with the facts that
  $Y^*$ is inclusion-minimal for $A$ in $\HP$ and that
  $\mathcal{H}\subseteq\HP$, $Y^*\in \mathcal{H}$ implies that
  $Y^*$ is also inclusion-minimal for $A$ in $\mathcal{H}$, i.e.\
  $Y^*=A_{\mathcal{H}}$. Analogous arguments imply $Y^*=B_{\mathcal{H}}$.
  Since $Y^*\cap B\ne\emptyset$ we conclude that $A\in
  \mathfrak{Y}$. Therefore, we have added a set $Y_A$ that satisfies both
  $A\subseteq Y_A$ and, by the arguments above, $Y_A\subsetneq Y^*$.
  Therefore, $Y^*$ is not inclusion-minimal for $A$ in $\HP$; a
  contradiction.

  In summary, $\HP$ is a hierarchy on $X$ such that
  Conditions~(i) and~(ii) in Thm.~\ref{thm:compatible} are satisfied for
  all $A,B\in\mathcal{P}$. Thus $\mathcal{P}$ is compatible with the
  refinement $\HP$ of $\mathcal{H}$.
\end{proof}

\begin{theorem}
  A hierarchy $\mathcal{H}$ and a partition $\mathcal{P}$ on $X$ are
    r-compatible if and only if no set $Y\in \mathcal{H}$ overlaps with
  two distinct sets $A,B\in \mathcal{P}$.
  \label{thm:compatible-refinement}
\end{theorem}
\begin{proof}
The \emph{only-if}-direction follows from
    Lemma~\ref{lem:refinement-implies-no-overlaps}.  Conversely, if no set
    $Y\in \mathcal{H}$ overlaps with two distinct sets
    $A,B\in \mathcal{P}$, then Lemma~\ref{lem:HP-properties}(2)
    and~\ref{lem:HP-compatible} imply that the hierarchy $\HP$ is
    compatible with $\mathcal{P}$.  By construction,
    $\mathcal{H}\subseteq \HP$ and thus, $\HP$ is a refinement of
    $\mathcal{H}$, which completes the proof.
\end{proof}

It is worth noting that the characterization of r-compatibility in
Thm.~\ref{thm:compatible-refinement} implies neither Property (i) nor
Property (ii) in Thm~\ref{thm:compatible}. As a counterexample to (i),
consider the hierarchy $\mathcal{H}$ on $X=\{a,a',b,b'\}$ that comprises in
addition to $X$ and the singletons only the set $Y=\{a,a',b\}$, and the
partition $\mathcal{P}=\{A,B\}$ with $A=\{a,a'\}$ and $B=\{b,b'\}$. We have
$A_{\mathcal{H}}=Y$ and $B_{\mathcal{H}}=X$. Clearly, $Y=A_{\mathcal{H}}$
overlaps $B$ and thus violates (i). On the other hand, it admits the
refinement $\mathcal{H}^*=\mathcal{H}\cup\{A\}$, which is compatible with
$\mathcal{P}$.  As a counterexample to (ii), consider $\mathcal{H}'$
comprising only $X$ and the singletons. Here, we have
$A_{\mathcal{H}'}=B_{\mathcal{H}'}=X$, while both refinements
$\mathcal{H}'\cup\{A\}$ and $\mathcal{H}'\cup\{B\}$ are compatible with
$\mathcal{P}$.

We continue with considering systems
$\mathfrak{P}=\{\mathcal{P}_1,\mathcal{P}_2,\dots,\mathcal{P}_k\}$ of
partitions of $X$ rather then single partitions.
By definition, $T$ (or equivalently the hierarchy
$\mc{H}(T)$) and each $\mathcal{P}_i$ are compatible if and only if
$\mathcal{P}_i=\mathcal{F}(T,H_i)$ for some subset $H_i\subseteq E(T)$,
$1\leq i\leq k$. In this case, we say that $T$ (equiv.\ $\mc{H}(T)$) and
$\mathfrak{P}$ are compatible. It is natural, then, to ask whether for a
given system of partitions $\mathfrak{P}$, there exists a
tree $T$ such that $\mathfrak{P}$ and $T$ are compatible.

\begin{proposition}
  Let $\mathcal{H}$ be a hierarchy on $X$ and
  $\mathfrak{P}=\{\mathcal{P}_1,\mathcal{P}_2,\dots,\mathcal{P}_k\}$ be a
  collection of partitions of $X$. The following two
  statements are
  equivalent
  \begin{enumerate}[noitemsep]
    \item[(1)] There is a refinement $\mathcal{H}^*$ of $\mathcal{H}$ that is
    compatible with $\mathfrak{P}$.
    \item[(2)] Each $\mathcal{P}_i\in \mathfrak{P}$ admits a compatible
    refinement $\mathcal{H}_i^*$ of $\mathcal{H}$ such that
    $\bigcup_{i=1}^k\mathcal{H}_i^*$ is a hierarchy.
  \end{enumerate}
  \label{prop:common-Psystems}
\end{proposition}
\begin{proof}
  Let $\mathcal{H}^*$ be a refinement of $\mathcal{H}$ that is compatible
  with $\mathfrak{P}$ and thus, with every $\mathcal{P}_i\in \mathfrak{P}$.
  Now, put $\mathcal{H}_i^* = \mathcal{H}^*$, $1\leq i\leq k$. Hence,
  $\mathcal{H}^* = \bigcup_{i=1}^k\mathcal{H}_i^*$ is a hierarchy
  that is    compatible with $\mathfrak{P}$.
  Conversely, if $\bigcup_{i=1}^k\mathcal{H}_i^*$ is a hierarchy, then it is,
  in particular, a refinement of every $\mc{H}_i^*$, $1\leq i \leq k$.  Now
  set $\mathcal{H}^* = \bigcup_{i=1}^k\mathcal{H}_i^*$.  By
  Prop.~\ref{prop:refinement}, $\mathcal{H}^*$ is compatible with with every
  $\mathcal{P}_i\in \mathfrak{P}$.
\end{proof}

\begin{figure}[t]
  \begin{center}
    \includegraphics[width=.6\textwidth]{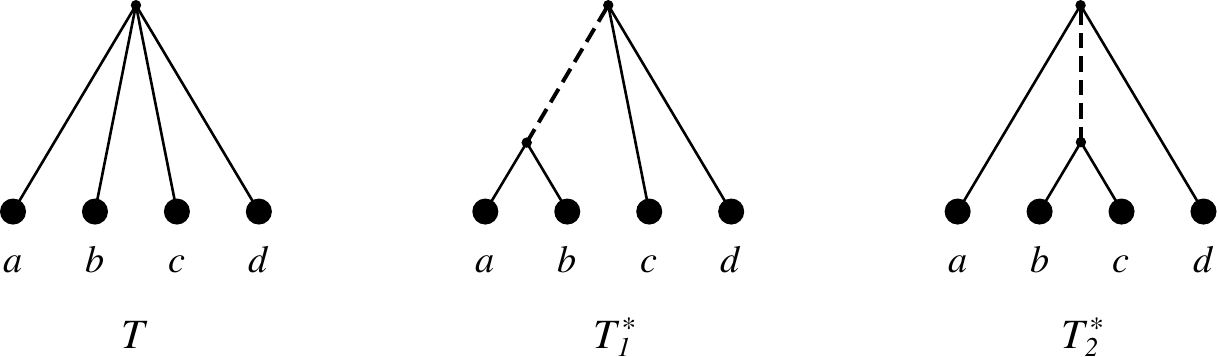}
  \end{center}
  \caption{The tree $T$ with leaf set $X=\{a,b,c,d\}$ is compatible with
    neither $\mathcal{P}_1 = \{\{a,b\},\{c,d\}\}$ nor
    $\mathcal{P}_2 = \{\{a,c\},\{b,d\}\}$. The refinements $T^*_1$ and
    $T^*_2$ are compatible with $P_1$ and $P_2$, respectively.  However,
    there is no refinement $\mathcal{H}^*(T)$ of $\mc{H}(T)$ (and thus of
    $\mc{H}(T^*_1)$ and $\mc{H}(T^*_2)$) that is compatible with both
    $\mathcal{P}_1$ and $\mathcal{P}_2$ as a consequence of
    Prop.\ \ref{prop:S-unroot} and the fact that the unrooted versions
    $\overline{T}_1^*$ and $\overline{T}_2^*$ of $T^*_1$ and $T^*_2$,
    resp., are determined by the splits $\{a,b\}|\{c,d\}$ and
    $\{a,d\}|\{b,c\}$, resp., which cannot be represented in a common
    unrooted tree (cf.\ ``Splits-Equivalence Theorem''
    \cite[Thm.~3.1.4]{Semple:03}).  }
  \label{fig:no-common-refinement}
\end{figure}

Prop.~\ref{prop:common-Psystems} immediately implies
  \begin{corollary}\label{cor:common-Psystems-Tree}
    There is a tree $T$ that is compatible with a collection
    $\mathfrak{P}=\{\mathcal{P}_1,\mathcal{P}_2,\dots,\mathcal{P}_k\}$ of
    partitions of $X$ if and only if, for every $i\in \{1,\dots,k\}$, there
    is a tree $T_i$ that is compatible with $\mathcal{P}_i$ such that
    $\bigcup_{i=1}^k\mathcal{H}(T_i)$ is a hierarchy.
  \end{corollary}
  \begin{proof}
    Since every tree on $X$ is a refinement of the star tree $T'$
    on $X$, every hierarchy  $\mathcal{H}^*$ on $X$ is a refinement of
    $\mathcal{H}(T')$.
    Hence, the existence of some hierarchy or, equivalently, some
    tree that is compatible with $\mathfrak{P}$
    is equivalent to the existence of a refinement of $\mathcal{H}^*$ of
    $\mathcal{H}(T')$ that is compatible with $\mathfrak{P}$.
  \end{proof}
As illustrated in Fig.~\ref{fig:notHierarchy}, $\mathcal{H}^*$ and
$\mathfrak{P}=\{\mathcal{P}_1,\mathcal{P}_2,\dots,\mathcal{P}_k\}$ might be
compatible, although there are refinements $\mathcal{H}_i^*$ of
$\mathcal{H}$ compatible with $P_i\in \mathfrak{P}$, whose union
$\bigcup_{i=1}^k \mathcal{H}_i^*$ does not form a hierarchy.
Fig.~\ref{fig:no-common-refinement}, furthermore, shows an example of
a partition system $\mathfrak{P}$ that is not compatible with any
refinement of $\mathcal{H}$.
We will show in Section~\ref{sec:compl-algos} that deciding whether or not
such a common refinement exists is an NP-complete problem.

The partitions of a set $X$ form a complete lattice
\cite[Sect.~4.9]{Birkhoff:67}.  The \emph{common refinement}
$\mathcal{P}_1 \wedge \mathcal{P}_2$ of two partitions $\mathcal{P}_1$ and
$\mathcal{P}_2$ of $X$ is
\begin{equation}
  \mathcal{P}_1 \wedge \mathcal{P}_2 \coloneqq
  \{ A_1\cap A_2 \mid A_1\in\mathcal{P}_1,\,A_2\in\mathcal{P}_2,\,A_1\cap
  A_2\ne\emptyset \}.
\end{equation}
The common refinement operation is associative and commutative. The common
refinement of an arbitrary system $\mathfrak{P}$ of partitions,
therefore, consists of all distinct sets
$P_x = \bigcap_{A\in \mathcal{P}\in \mathfrak{P},\,x\in A} A$.  The
following results shows that the common refinement of partitions that are
compatible with $T$ is compatible with $T$ as well.
\begin{proposition}
  Let $T$ be a tree with leaf set $X$ and
  $\mathfrak{P}=\{\mathcal{P}_1,\mathcal{P}_2,\dots,\mathcal{P}_k\}$ a
  collection of partitions on $X$ that are all compatible with
  $T$. Then, $\bigwedge_{i=1}^k \mathcal{P}_i$ is compatible with $T$.
  \label{prop:wedge}
\end{proposition}
\begin{proof}
  Let $T$ be a tree with leaf set $X$.  We show first that for all subsets
  $H_1,H_2\in E(T)$ it holds that
  \begin{equation}
    \mathcal{F}(T,H_1\cup H_2) = \mathcal{F}(T,H_1)\wedge\mathcal{F}(T,H_2).
  \end{equation}
  To see this, let $A\in \mathcal{F}(T,H_1)\wedge\mathcal{F}(T,H_2)$. For all
  $a,a'\in A$ the following statements are equivalent
  \begin{enumerate}
    \item $a,a'\in A = A_1\cap A_2$ for some
    $A_1\in\mathcal{F}(T,H_1),\, A_2\in\mathcal{F}(T,H_2)$,
    \item there is no separating edge in $H_1$ and in $H_2$ that is on the path
    between $a$ and $a'$ in $T$,
    \item there is no separating edge in $H_1\cup H_2$ that is on the path
    between
    $a$ and $a'$ in $T$, and
    \item $a,a'\in B \in \mathcal{F}(T,H_1\cup H_2)$.
  \end{enumerate}
  Hence, $A\subseteq B\in \mathcal{F}(T,H_1\cup H_2)$. Similarly, the latter
  equivalent statements hold for all
  $a,a'\in B \in \mathcal{F}(T,H_1\cup H_2)$ and thus, $A=B$.

  Now assume that $\mathcal{P}_1,\mathcal{P}_2,\dots,\mathcal{P}_k$ are all
  compatible with $T$.  Hence, for each $\mathcal{P}_i$ there is a set
  $H_i\subseteq E(T)$ such that $\mathcal{P}_i = \mathcal{F}(T,H_i)$. By the
  latter arguments and since $\wedge$ is commutative and associative, we can
  conclude that
  $\bigwedge_{i=1}^k \mathcal{P}_i = \bigwedge_{i=1}^k \mathcal{F}(T,H_i) =
  \mathcal{F}(T,\cup_{i=1}^k H_i)$ and thus,
  $\bigwedge_{i=1}^k \mathcal{P}_i$ is compatible with $T$.
\end{proof}

The converse of Prop.~\ref{prop:wedge} is not true in general.  As an
example, consider the tree $T$ as shown in Fig.~\ref{fig:no-common-refinement}
and the two partitions
$\mathcal{P}_1 = \{\{a,b\},\{c,d\}\}$ and
$\mathcal{P}_2 = \{\{a,c\},\{b,d\}\}$. Both $\mathcal{P}_1$ and
$\mathcal{P}_2$ are not compatible with $T$, however, their common
refinement
$\mathcal{P}_1\wedge \mathcal{P}_2 =\{\{a\}, \{b\}, \{c\}, \{d\}\} =
\mathcal{F}(T,E(T))$ is.

The \emph{refinement supremum} or \emph{join}
$\mathcal{P}_1\vee\mathcal{P}_2$ is obtained by recursively unifying any
two sets $A_1,A'_1\in\mathcal{P}_1$ whenever there is $A_2\in\mathcal{P}_2$
such that $A_1\cap A_2\ne\emptyset$ and $A_1'\cap A_2\ne\emptyset$.  The
analogue to Prop.~\ref{prop:wedge} does not hold for the refinement
supremum.  To see this, consider the tree $T$ on $X=\{a,b,c,d\}$ with
  hierarchy
  $\mathcal{H}(T) = X\cup \{\{x\}\mid x\in X\} \cup \{\{a,b\},\{a',b'\}\}$
and the partitions
$\mathcal{P}_1=\{\{a,a'\},\{b\},\{b'\}\}$ and
$\mathcal{P}_2=\{\{a\},\{a'\},\{b,b'\}\}$.  Both $\mathcal{P}_1$ and
$\mathcal{P}_2$ are compatible with $T$ (just define the edges incident to
the $x$ for respective singletons $\{x\}$ as separating edges).  However,
$\mathcal{P}_1 \vee \mathcal{P}_2 = \{A\coloneqq\{a,a'\},
B\coloneqq\{b,b'\}\}$ is not compatible since, for the hierarchy
$\mathcal{H}$ corresponding to $T$, we have
$A_{\mathcal{H}}=B_{\mathcal{H}}$; a contradiction to Condition~(ii) in
Thm.~\ref{thm:compatible}. In \cite{Tochon:19}, a notion of local
comparability of partitions is considered:
$\mathcal{P}_1\simeq\mathcal{P}_2$ iff
$A_1\cap A_2\in \{\emptyset,A_1,A_2\}$ for all $A_1\in\mathcal{P}_1$ and
$A_2\in\mathcal{P}_2$. The example above satisfies
$\mathcal{P}_1\simeq\mathcal{P}_2$. Hence, local comparability of
partitions $\mathcal{P}_1$ and $\mathcal{P}_2$ compatible with a given
hierarchy $\mathcal{H}$ is also not sufficient to imply compatibility of
$\mathcal{P}_1 \vee \mathcal{P}_2$ and $\mathcal{H}$.

\section{Compatibility of Partitions with Split Systems and Unrooted Trees}
\label{sec:splits}

Throughout this section, we will assume that all unrooted trees are
phylogenetic as well and have at least three leaves. In particular,
therefore, they have at least one inner vertex. Not surprisingly, there is a
very close connection between the case of rooted and unrooted phylogenetic
trees.
\begin{proposition}
  Let $\overline T$ be an unrooted tree with leaf set $X$ and let
  $\mathcal{P}$ be a partition of $X$. Then $\mathcal{P}$ and
  $\overline{T}$ are compatible if and only if $\mathcal{P}$ and the rooted
  tree $T$ are compatible, where $T$ is obtained by rooting $\overline T$
  at an arbitrary inner vertex.
  \label{prop:S-root}
\end{proposition}
\begin{proof}
  Note that rooting $\overline T$ at an arbitrary inner vertex results in a
  \emph{phylogenetic} rooted tree $T$, since $\overline T$ does not contain
  vertices of degree two. The equivalence now follows immediately from the
  definition and the fact that, viewed as pair of a set vertices and a set
  of edges, $T=\overline T$ and therefore,
  $\mathcal{P}=\mathcal{F}(\overline T,H) = \mathcal{F}( T,H)$ for some
  subset $H\subseteq E(\overline T)$.
\end{proof}

In fact, it is not necessary to root $\overline{T}$ at an inner vertex, we
may as well place the root as a subdivision of any edge. This allows us to
connect unrooted trees directly to hierarchies:
\begin{proposition}
  Let $\mathcal{H}$ be a hierarchy with corresponding rooted tree $T$ with
  leaf set $X$ and let $\mathcal{P}$ be a partition of $X$. Then
  $\mathcal{H}$ and $\mathcal{P}$ are compatible if and only if the
  unrooted tree $\overline T$ obtained from $T$ (by suppressing a
  possible degree-two root) is compatible with $\mathcal{P}$.
  \label{prop:S-unroot}
\end{proposition}
\begin{proof}
  If the root $\rho$ of $T$ has degree greater than two, then the tree
  $\overline T$ is phylogenetic and thus, can we apply the same arguments
  as in the proof of Prop.~\ref{prop:S-root} to establish the equivalence.
  If $\rho$ has degree $2$, then the tree $\overline T$ is not phylogenetic
  and both edges $e_1=\{\rho,v_1\}$ and $e_2=\{\rho,v_2\}$ that are
  incident to $\rho$ define the same split $S_{e_1}=S_{e_2}$. Since
  $\mathfrak{S}(\overline T)$ contains a split only once, the unique
  phylogenetic tree $\overline T$ defined by the split system
  $\mathfrak{S}(\overline T)$ is obtained from $\overline T$ by suppressing
  the root $\rho$, i.e., $e^*\coloneqq \{v_1,v_2\}\in E(\overline T)$. Now
  let $H\subseteq E(T)$ be such that $\mc{F}(T,H)=\mc{P}$. If $e_1$ or
  $e_2$ is contained in $H$, we add the edge $e^*$ to
  $H\setminus\{e_1,e_2\}$ to obtain the set $H^*\subseteq E(\overline T)$
  and, otherwise, we put $H^*=H$.  It is now easy to see that
  $\mathcal{F}(T,H)=\mathcal{F}(\overline T,H^*)$ and thus, $\overline T$
  and $\mc{P}$ are compatible.
\end{proof}

As outlined in Section \ref{sec:prelim}, every unrooted phylogenetic tree
$\overline{T}$ is determined by its split system
$\mathfrak{S}(\overline{T}) = \{\mathcal{S}_e \colon e\in E(\overline T)\}$
and for a \emph{tree-like} split system $\mathfrak{S}$ there is a (unique)
unrooted tree $\overline{T}$ with
$\mathfrak{S}(\overline{T})=\mathfrak{S}$.  Since unrooted trees are so
intimately related with split systems, it is interesting, therefore, to ask
whether the compatibility of rooted trees or hierarchies and partitions can
also be expressed in an interesting way in terms of split systems.
\begin{corollary}
  For every partition $\mathcal{P}$ of a non-empty set $X$, the split
  system
  $\mathfrak{S}_{\mathcal{P}}^*\coloneqq \mathfrak{S}_{\mathcal{P}}\cup\{
  x|(X\setminus\{x\}) \colon x\in X \})$ with
  \begin{equation}
    \mathfrak{S}_{\mathcal{P}} \coloneqq \left\{A|(X\setminus A) \colon
    A\in\mathcal{P}\right\}
  \end{equation}
  is always tree-like and compatible with $\mathcal{P}$.
  \label{cor:specialS}
\end{corollary}
\begin{proof}
  It is easy to see that the hierarchy $\mathcal{H}_{\mathcal{P}}$ (as
  specified in Lemma \ref{lem:specialH}) yields a rooted tree $T$ for which
  the unrooted version $\overline{T}$ (by suppressing a possible degree-two
  root of $T$) satisfies
  $\mathfrak{S}(\overline{T}) = \mathfrak{S}_{\mathcal{P}}^*$.  Now apply
  Prop.\ \ref{prop:S-unroot}.
\end{proof}

\begin{figure}[t]
  \begin{center}
    \includegraphics[width=.25\textwidth]{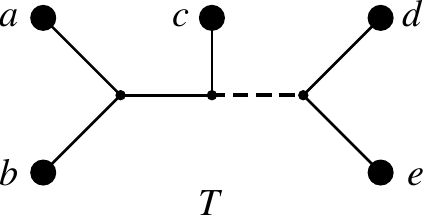}
  \end{center}
  \caption{The unrooted tree $T$ and the partition
    $\mathcal{P}=\{\{a,b,c\},\{d,e\}\}$ are compatible. However,
    $\mathfrak{S}(\overline{T})\neq \mathfrak{S}_{\mathcal{P}}^*$, because
    for $\mathcal{S} =\{a,b\}|\{c,d,e\}\in \mathfrak{S}(\overline{T})$ we
    have
    $\mathcal{S}\notin \mathfrak{S}_{\mathcal{P}} =
    \{\{a,b,c\}|\{d,e\}\}$.}
  \label{fig:splits}
\end{figure}

Compatibility of a partition $\mathcal{P}$ with a split system
$\mathfrak{S}(\overline{T})$ of some tree, however, does not imply that
$\mathfrak{S}(\overline{T})=\mathfrak{S}_{\mathcal{P}}^*$, see Fig.\
\ref{fig:splits}.  Suppose that $\overline{T}$ and $\mathcal{P}$ are
compatible, i.e., that $\mathcal{P}=\mathcal{F}(\overline{T},H)$.  The set
$H$ of separating edges then corresponds to the set
\begin{equation*}
  \mathfrak{H}_{H,\overline{T}}\coloneqq
  \{\mathcal{S}_e \colon e\in H\}\subseteq\mathfrak{S}(\overline{T})\,.
\end{equation*}
of splits.  Then we have either
\begin{equation}
  B\subseteq A \text{ or } B\subseteq X\setminus A \qquad \text{for every }
  A|(X\setminus A)\in\mathfrak{H}_{H,\overline{T}}
  \text{ and every } B\in\mathcal{P}
  \label{eq:H-no-set-cut}
\end{equation}
since none of the edges $e\in H$ separates two vertices of
$B$. Furthermore, for any two distinct sets $A,A'\in\mathcal{P}$ there is a
split $B|B'\in\mathfrak{H}_{H,\overline{T}}$ such that, w.l.o.g.,
$A\subseteq B$ and $A'\subseteq B'$ because there must be an edge in $H$
separating $A$ and $A'$ in $\overline{T}$. Taken together,
therefore, we
observe that every set $B\in\mathcal{P}$ satisfies
\begin{equation}
  B =  \bigcap \left\{ A\in 2^X\,:\, A|(X\setminus A)\in
  \mathfrak{H}_{H,\overline{T}},\, B\subseteq A \right\}.
  \label{eq:splits}
\end{equation}
Conversely, suppose that an arbitrary split system $\mathfrak{H}$ satisfies
Eqs.~\eqref{eq:H-no-set-cut} and~\eqref{eq:splits} (replace
$\mathfrak{H}_{H,\overline{T}}$ by $\mathfrak{H}$ in the equations).  By
Eq.~\eqref{eq:splits}, for every $B'\in\mathcal{P}$ with $B\ne B'$ and thus
$B\cap B'=\emptyset$, there is a split $A|(X\setminus A)$ such that
$B\subseteq A$ and $B'\subseteq X\setminus A$ and thus there is an edge
$e_{BB'}\in E(\overline T))$ that separates $B$ and $B'$ in
$\overline{T}$. Add all these edges to the set $H$.
Eq.~\eqref{eq:H-no-set-cut} ensures that no two elements in $B$ are
separated by an edge in $H$.  Thus, $B\in\mathcal{F}(\overline{T},H)$ for
every $B\in\mathcal{P}$. Hence, $\mathcal{P}$ and $\overline{T}$ are
compatible if and only if $\mathfrak{H}\subseteq\mathfrak{S}(\overline{T})$
satisfies Eqs.~\eqref{eq:H-no-set-cut} and~\eqref{eq:splits}. Recall that
splits on $X$ are partitions of $X$ and that the common refinement of
a set $\mathfrak{P}$ of partitions of $X$ is the partition
$\bigwedge\mathfrak{P}$ whose sets are the intersections
$B_x\coloneqq \bigcap\{B\in\mathcal{P}\in\mathfrak{P} \text{ s.t.\ } x\in
B\}$ of sets appearing in any of the partitions in the system
$\mathfrak{P}$ that have a least one point $x\in X$ in common. Thus,
$\mathfrak{H}\subseteq\mathfrak{S}(\overline{T})$ satisfies
Eqs.~\eqref{eq:H-no-set-cut} and~\eqref{eq:splits} if and only if
$\mathcal{P}=\bigwedge\mathfrak{H}$. We summarize this discussion as
\begin{theorem}
  Let $\mathcal{P}$ be a partition and $\mathfrak{S}$ be a tree-like split
  system. Then $\mathcal{P}$ and $\mathfrak{S}$ are compatible if and only
  if there is subset $\mathfrak{H}\subseteq \mathfrak{S}$ such that
  $\mathcal{P}$ is the common refinement of $\mathfrak{H}$.  In this case,
  $\mathcal{P} = \mathcal{F}(\overline{T},H)$ for the tree $\overline{T}$
  with $\mathfrak{S}(\overline{T}) = \mathfrak{S}$ and
  $H = \{e\mid e\in E(\overline{T}), S_e\in \mathfrak{H}\}$.
  \label{thm:splitcomp}
\end{theorem}
Since every refinement $\overline{T}^*$ of a tree $\overline{T}$
corresponds to a tree-like split system $\mathfrak{S}(\overline{T}^*)$
that satisfies
$\mathfrak{S}(\overline{T})\subseteq\mathfrak{S}(\overline{T}^*)$, we
immediately obtain a characterization of tree-like split systems that admit
a refinement that is compatible with a partition $\mathcal{P}$.
\begin{corollary}
  Let $\mathcal{P}$ be a partition and $\mathfrak{S}(\overline{T})$ be the
  split system associated with a tree $\overline{T}$. Then $\mathcal{P}$ is
  compatible with a refinement of $\overline{T}$ if and only if there is a
  set of splits $\mathfrak{H}$ such that (i) $\mathcal{P}$ is the common
  refinement of $\mathfrak{H}$ and (ii)
  $\mathfrak{S}(\overline{T})\cup\mathfrak{H}$ is tree-like.
  \label{cor:splitrefcomp}
\end{corollary}
The characterizations in Thm.~\ref{thm:splitcomp} and
Cor.~\ref{cor:splitrefcomp} are not constructive, i.e., they do not provide
recipes to construct $\mathfrak{H}$.  Clearly, we can directly employ
Prop.~\ref{prop:S-root} and the linear-time algorithm provided in
Section~\ref{sec:compl-algos} to check whether a split system and a partition
are
compatible or not. Nevertheless, we provide the following two lemmas to
provide a further constructive characterization that makes use of a
step-wise decomposition and might be of further theoretical interest.

\begin{definition}[{\cite[Def. 6.1.1]{Semple:03}}]
  Let $\overline{T}$ be an unrooted phylogenetic tree with leaf set $X$ and
  corresponding split system $\mathfrak{S}(\overline{T})$. Then the
  restriction $\overline{T}_{|Y}$ of $\overline{T}$ to a non-empty subset
  $Y\subseteq X$ of leaves is the tree for which
  \begin{equation*}
    \mathfrak{S}(\overline{T}_{|Y}) = \{ (A\cap Y)|(Y\setminus A) \,:\,
    A|(X\setminus A)\in\mathfrak{S}(\overline T),\,  A\cap
    Y\ne\emptyset,\, Y\setminus A\ne \emptyset\,\}.
  \end{equation*}
\end{definition}

Somewhat surprisingly, it suffices to check whether there is a single
element $A\in\mathcal{P}$ such that $\mathcal{P}\setminus\{A\}$ and
$\overline{T}_{|X\setminus A}$ are compatible, to determine whether
$\mathcal{P}$ and $\overline{T}$ are compatible as shown in the next
\begin{lemma}
  Let $\overline{T}$ be an unrooted tree with leaf set $X$
  and let $\mathcal{P}$ be a partition of $X$. Then
  $\mathcal{P}$ and $\overline{T}$ are compatible if and only if
  $|\mathcal{P}|=1$, or there is a set $A\in\mathcal{P}$ such that (i)
  $A| (X\setminus A)\in\mathfrak{S}(\overline{T})$ and (ii)
  $\mathcal{P}\setminus\{A\}$ and $\overline{T}_{|X\setminus A}$
  are compatible.
  \label{lem:splitcompat}
\end{lemma}
\begin{proof}
  Clearly, every tree on $X$ is compatible with the partition
  $\mathcal{P} =\{X\}$ since
  $\{X\}=\mathcal{F}(\overline{T},\emptyset)$. Thus, assume
  $|\mathcal{P}|>1$ in the following.  First suppose that $\overline{T}$
  and $\mathcal{P}$ are compatible.  We first show that there is a set
  $A\in\mathcal{P}$ such that $A|(X\setminus A)\in\mathfrak{S}(\overline{T})$.
  Since
  $\overline{T}$ and $\mathcal{P}$ are compatible, there is an edge set
  $H\subseteq E(\overline{T})$ such that
  $\mathcal{F}(\overline{T},H)=\mathcal{P}$
  and a corresponding set of splits
  $\mathfrak{H} \coloneqq
  \mathfrak{H}_{H,\overline{T}}\subseteq\mathfrak{S}(\overline{T})$.
  Since $|\mathcal{P}|>1$ by
  assumption, both sets $H$ and $\mathfrak{H}$ are non-empty. Therefore, we
  can pick an arbitrary split $S_1|(X\setminus S_1)\in\mathfrak{H}$. By
  Eq.~\eqref{eq:H-no-set-cut}, every $A\in\mathcal{P}$
  satisfies either $A\subseteq S_1$
  or $A\subseteq X\setminus S_1$.
  In particular, there must be some $A\in\mathcal{P}$ with $A\subseteq S_1$.
  If there is no other split
  $S_2|(X\setminus S_2)\in\mathfrak{H}$ with $S_2\subsetneq S_1$, then
  $S_1\in\mathcal{P}$  and
  we are done. Otherwise consider the split
  $S_2|(X\setminus S_2)\in\mathfrak{H}$ with $S_2\subsetneq S_1$. Then either
  there is another split
  $S_3|(X\setminus S_3)\in\mathfrak{H}$ with $S_3\subsetneq S_2$ or
  $S_2\in\mathcal{P}$. Since $\mathfrak{H}$ is finite, we eventually reach
  a split $S_j|(X\setminus S_j)\in\mathcal{H}$ with $S_j\in\mathcal{P}$,
  i.e., we can choose $A=S_j$ in Condition (i). Now let $e_A\in H$ be the
  edge in $\overline{T}$ with $S_e = A|(X\setminus A)$.  The
  restriction $\overline{T}_{|Y}$ of $\overline{T}$ to $Y=X\setminus A$ is
  therefore simply the connected component $\overline{T}'$ of
  $\overline{T}-e_A$ that does not intersect $A$. Thus
  $\mathcal{F}(\overline{T}_{|X\setminus A},H\setminus\{e_A\})=
  \mathcal{P}\setminus\{A\}$, i.e., Condition (ii) is satisfied.

  Conversely, suppose there is an $A\in\mathcal{P}$ such that
  $\mathcal{P}\setminus\{A\}$ and $\overline{T}_{|X\setminus A}$ is
  compatible and $A|(X\setminus A)$ is a split corresponding to an edge in
  $\overline{T}$. Then there is an edge set
  $H'\subseteq E(\overline{T}_{|X\setminus A})$ such that
  $\mathcal{F}(\overline{T}_{|X\setminus A},H')=\mathcal{P}'=
  \mathcal{P}\setminus\{A\}$. Since $A|(X\setminus
  A)\in\mathfrak{S}(\overline{T})$, there
  is an edge $e_a$ in $E(\overline{T})\setminus E(\overline{T}_{|X\setminus A})$
  connecting the subtrees $\overline{T}_{|X\setminus A}$ and
  $\overline{T}_{|A}$. In particular $e_A\notin H'$. Therefore, we have
  $\mathcal{F}(\overline{T},H'\cup\{e_A\})=
  \mathcal{F}(\overline{T}_{|X\setminus A},H')\cup\{A\}=
  \mathcal{P}'\cup\{A\}=\mathcal{P}$, i.e., $\overline{T}$ and
  $\mathcal{P}$ are compatible.
\end{proof}

\begin{lemma}
  Let $\mathcal{P}$ be a partition of $X$, and $\overline{T}$ an unrooted
  tree with leaf set $X$. Then
  $\overline{T}$ admits a refinement $\overline{T}^*$ compatible with
  $\mathcal{P}$ if and only if $|\mathcal{P}|=1$, or there is an
  $A\in\mathcal{P}$ such that (i) for every $B_1|B_2
  \in\mathfrak{S}(\overline{T})$ we
  have at least one of $A\subseteq B_1$, $A\subseteq B_2$,
  $B_1\subseteq A$, or $B_2\subseteq A$ and (ii) the restriction
  $\overline{T}_{|X\setminus A}$ admits a refinement
  $\overline{T}^*_{|X\setminus A}$ that is compatible with
  $\mathcal{P}\setminus\{A\}$.
  \label{lem:splitrefine}
\end{lemma}
\begin{proof}
  For brevity, we write $\mathfrak{T}\coloneqq \mathfrak{S}(\overline{T})$
  and $\mathfrak{T}^*\coloneqq \mathfrak{S}(\overline{T}^*)$ for the split
  systems of the two trees $T$ and $T^*$.  Clearly, every tree on $X$ is
  compatible with the partition $\mathcal{P} =\{X\}$ since
  $\{X\}=\mathcal{F}(\overline{T},\emptyset)$. Thus, assume
  $|\mathcal{P}|>1$ in the following.  Suppose the refinement
  $\overline T^*$ (with corresponding split system $\mathfrak{T}^*$) of
  $\overline T$ is compatible with $\mathcal{P}$. By
  Lemma~\ref{lem:splitcompat}, there is an $A\in\mathcal{P}$ such that
  $A|(X\setminus A)$ is a split in $\mathfrak{T}^*$ and
  $\overline{T}^*_{|X \setminus A}$ is compatible with
  $\mathcal{P}\setminus\{A\}$. Clearly, $\overline{T}^*_{|X \setminus A}$
  is a refinement of $\overline{T}_{|X \setminus A}$, i.e.,
  $\overline{T}_{|X \setminus A}$ admits a refinement that is compatible
  with $\mathcal{P}\setminus\{A\}$. Since $\mathfrak{T}^*$ identifies the
  tree $\overline{T}^*$, it is in particular a tree-like split
  system.  Since $\overline{T}^*$ is a refinement of $\overline{T}$, we
  have $\mathfrak{T}\subseteq\mathfrak{T}^*$. In particular, therefore
  $A|(X\setminus A)$ and every split $B_1|B_2\in\mathfrak{T}$ must have at
  least one empty intersection $A\cap B_1$, $A\cap B_2$,
  $(X\setminus A)\cap B_1$ or $(X\setminus A)\cap B_2$. Depending on which
  of the four intersections is empty, we have one of the following
  situations $A\subseteq B_2$, $A\subseteq B_1$, $B_1\subseteq A$ or
  $B_2\subseteq A$, respectively.

  Conversely, suppose there is an $A\in\mathcal{P}$ satisfying conditions
  (i) and (ii). Then
  $\mathfrak{T}'\coloneqq \mathfrak{T}\cup\{ A|(X\setminus A)\}$ is a
  tree-like split system because $\mathfrak{T}$ has this property and, for
  any $B_1|B_2\in \mathfrak{T}\setminus \{A|(X\setminus A)\}$, the
  alternatives in (i) amount to $A\cap B_2=\emptyset$,
  $A\cap B_1=\emptyset$, $(X\setminus A)\cap B_1=\emptyset$, or
  $(X\setminus A)\cap B_2=\emptyset$, respectively. Moreover,
  $\mathfrak{T}$ and thus $\mathfrak{T}'$ contains the singleton splits
  $\{x\}|(X\setminus\{x\})$ for all $x\in X$.  The split system
  $\mathfrak{T}'$ therefore defines a refinement $\overline{T}'$ of
  $\overline{T}$. Furthermore, we have
  $\overline{T}_{|X\setminus A}=\overline{T}'_{|X\setminus A}$ and
  $\overline{T}_{|A}=\overline{T}'_{|A}$ since either
  $\overline{T}=\overline{T}'$ or the difference between $\overline{T}$ and
  $\overline{T}'$ is only the expansion or contraction of the edge $e_A$
  identified by the additional split $A|(X\setminus A)$. By condition~(ii),
  there is a refinement $\overline{T}^*_{|X\setminus A}$ of the restriction
  $\overline{T}_{|X\setminus A}=\overline{T}'_{|X\setminus A}$ that is
  compatible with the partition $\mathcal{P}\setminus\{A\}$ of
  $X\setminus A$.  Thus there is also a refinement ${\overline{T}'}^*$ of
  $\overline{T}$ such that the restriction
  ${\overline{T}'}^*_{|X\setminus A}=\overline{T}^*_{|X\setminus A}$ is
  compatible with $\mathcal{P}\setminus\{A\}$. Let $H^*$ be the
  corresponding set of separating edges.  Then
  $\mathcal{F}({\overline{T}'}^*,H^*\cup\{e_A\})=
  \mathcal{F}(\overline{T}^*_{|X\setminus\{A\}},H^*)\cup\{A\} =
  \mathcal{P}$. Thus $\overline{T}$ and $\mathcal{P}$ are compatible.
\end{proof}

Lemma~\ref{lem:splitcompat} and Lemma~\ref{lem:splitrefine} are associated
with a simple algorithmic intuition. Among the sets $A\in\mathcal{P}$, at
least one corresponds (at least in a refinement) to a connected component
in the forest obtained by deletion of a single edge, and thus to a split
$A|(X\setminus A)$ that is either already contained in
$\mathfrak{S}(\overline{T})$ or that can be used to refine the tree
$\overline{T}$. This immediately yields an algorithm on the split systems
that, in each step, finds a set $A\in\mathcal{P}$ that satisfies condition
(i) and then proceeds to checking the restriction to $X\setminus A$. While
the formulation in terms of tree-like split systems is of some theoretical
interest, it seems to be of little practical use compared to the
linear-time algorithms described in the following section.

\section{Algorithms and Complexity}
\label{sec:compl-algos}

In the following, we will first derive results for the complexity of
  checking whether $\mc{P}$ and $T$ are r-compatible and the construction
  of the edge coloring $\gamma_{T,\mc{P}}$ which we then use for the
  special case of compatibility of $\mc{P}$ and $T$.  Finally, we
  investigate the complexity of finding a refinement that is compatible
  with a system $\mathfrak{P}$ of partitions.  In view of
  Prop.~\ref{prop:S-unroot}, we will assume that $T$ is a rooted tree in
  this section unless explicitly stated otherwise.

Recall that, for a tree $T$ on $X$ and a partition $\mc{P}$ of $X$,
  the map $\gamma_{T,\mc{P}}$ assigns to $e\in E(T)$ as ``colors'' all sets
  $A\in\mc{P}$ for which $e$ lies on a path connecting two elements
  $x,x'\in A$.

\begin{fact}
  Let $T$ be a tree on $X$ and $\mathcal{P}$ a partition of $X$,
  $v\ne\rho_T$, and $e=\{\parent(v),v\}\in E(T)$. Then
  \begin{equation*}
    A\in \gamma_{T,\mathcal{P}}(e) \iff
    A\cap L(T(v))\ne\emptyset \text{ and } A\setminus L(T(v))\ne\emptyset.
  \end{equation*}
  \label{obs:edge-color-i}
\end{fact}

\begin{lemma}
  Let $\mathcal{H}$ be a hierarchy on $X$, $T$ the corresponding tree on $X$,
  and $\mathcal{P}$ a partition of $X$. Moreover, let $Y\in
  \mc{H}$.
  Then, $Y$ overlaps with two distinct sets
  $A,B\in\mathcal{P}$ if and only if $A,B\in
  \gamma_{T,\mathcal{P}}(\{\parent(\lca_T(Y)), \lca_T(Y)\})$ with $A\ne B$.
  \label{lem:over-iff-color-ij}
\end{lemma}
\begin{proof}
  Since $Y\in \mc{H}$, there is a unique vertex $v\in V(T)$ with $Y=L(T(v))$
  and thus, $v=\lca_T(Y)$.
  First assume that $Y\in \mathcal{H}$ overlaps with two distinct sets
  $A,B\in\mathcal{P}$.
  Thus, we have $C\cap L(T(v))\ne\emptyset$ and $C\setminus
    L(T(v))\ne\emptyset$ for $C\in\{A,B\}$.
  By Obs.~\ref{obs:edge-color-i}, this implies $A,B\in
  \gamma_{T,\mathcal{P}}(\{\parent(v), v\})$.
  For the converse, assume that $A,B\in
  \gamma_{T,\mathcal{P}}(\{\parent(v), v\})$ with $A\ne B$.
  By Obs.~\ref{obs:edge-color-i}, we have $C\cap Y\ne\emptyset$ and
    $C\setminus Y\ne\emptyset$ for $C\in\{A,B\}$.
  Since, in addition, $A$ and $B$ are disjoint, we have $Y\setminus
    C\ne\emptyset$ for $C\in\{A,B\}$.
  In summary, $Y$ overlaps with the two distinct sets $A$ and $B$.
\end{proof}

Lemma~\ref{lem:over-iff-color-ij} together with
Thm.~\ref{thm:compatible-refinement} immediately implies
\begin{proposition}
  Let $T$ be a tree on $X$ and $\mathcal{P}$ a partition of $X$.
  $\mc{P}$ and $T$ are r-compatible if and only if
  $|\gamma_{T,\mathcal{P}}(e)| \le 1$ for every $e\in E(T)$.
  \label{prop:at-most-one-color}
\end{proposition}

To find a refinement of a hierarchy $\mc{H}$ that is compatible with
$\mathcal{P}$, it is crucial to know the set
$\mathfrak{Y}(\mathcal{H},\mathcal{P})$, and, in particular, the sets
$A_{\mc{H}}$ for the $A\in\mathfrak{Y}(\mathcal{H},\mathcal{P})$.
However, an explicit construction of the latter
is not needed since the property of $Y\in \mc{H}$ being
equal to $A_{\mc{H}}$ for some $A\in\mathfrak{Y}(\mathcal{H},\mathcal{P})$
or not is entirely determined by the
colored edges incident to $\lca_T(Y)$.
\begin{lemma}
  Let $\mathcal{H}$ be a hierarchy on $X$, $T$ the corresponding tree on $X$,
  $\mathcal{P}$ a partition of $X$, and $Y\in
  \mathcal{H}$. Then, $Y=A_{\mc{H}}$ for some
  $A\in\mathfrak{Y}(\mathcal{H},\mathcal{P})$
  if and only if the following two conditions are
  satisfied for $u\coloneqq \lca_T(Y)$:
  \begin{enumerate}
    \item[(a')] $A\in \gamma_{T,\mathcal{P}}(\{u,v\})$ for some
    $v\in\child_T(u)$ and either $u=\rho_T$ or $A\notin
    \gamma_{T,\mathcal{P}}(\{\parent_{T}(u),u\})$.
    \item[(b')] $B\in \gamma_{T,\mathcal{P}}(\{u,v'\})$ for some
    $v'\in\child_T(u)$ and some color $B\neq A$.
  \end{enumerate}
  \label{lem:Y-localColoring}
\end{lemma}
\begin{proof}
  Let $Y=A_{\mc{H}}$ for some
  $A\in\mathfrak{Y}(\mathcal{H},\mathcal{P})\subseteq \mathcal{P}$.
  By definition, this is, if and only if,
  (a) $Y=A_{\mc{H}}$ with $A\in\mc{P}$ and (b) there is some
  $B\in\mc{P}\setminus\{A\}$
  satisfying $B\cap Y\ne\emptyset$ and $Y\subseteq B_{\mathcal{H}}$.
  In particular, $u=\lca_T(Y)$ is an inner vertex in this case.
  Thus, we have $L(T(u)) = Y=A_{{\mc{H}}}$.
  The definition of $\gamma_{T,\mathcal{P}}$ directly implies Condition (a').
  Now, let $B\in\mathcal{P}\setminus\{A\}$
  such that $B\cap Y\ne\emptyset$ and $Y\subseteq B_{\mathcal{H}}$.
  Hence, $B\cap L(T(u))\ne\emptyset$,
  and thus, there must be a child $v'\in\child_T(u)$ with $B\cap
  L(T(v'))\ne\emptyset$. However, since $L(T(v'))\subsetneq
  L(T(u))=Y\subseteq B_{\mathcal{H}}$ and since $B_{\mathcal{H}}$ is
  inclusion-minimal for $B$, we also have $B\setminus
  L(T(v'))\ne\emptyset$. Taken together, we obtain
  $B\in \gamma_{T,\mathcal{P}}(\{u,v'\})$ by definition of
  $\gamma_{T,\mathcal{P}}$ and thus Conditions (b').

  Now assume that Conditions (a') and (b') are satisfied.
  Let $A\in \gamma_{T,\mathcal{P}}(\{u,v\})$ for some  $v\in\child_T(u)$.
  Hence, $Y\subseteq A_{\mc{H}}$.  Moreover,
  the two possible cases  $u=\rho_T$ or $A\notin
  \gamma_{T,\mathcal{P}}(\{\parent_{T}(u),u\})$
  imply that
  there must be a second edge $\{u,v''\}$ for some
  $v''\in\child_T(u)\setminus \{v\}$
  that is colored with $A$ due to the definition of $\gamma_{T,\mathcal{P}}$
  and the fact that $u =\lca_T(Y)$. Therefore, $u= \lca_T(A_{\mc H}) =
  \lca_T(Y)$.
  This together with $Y\in \mathcal{H}$
  and $Y\subseteq A_{\mc{H}}$ implies Condition (a) $Y=A_{\mc{H}}$.
  Condition (b') implies that
  $B\cap L(T(v'))\ne\emptyset$ for some $v'\in\child_T(u)$ (and thus
  $B\cap L(T(u))\ne\emptyset$)
  and $B\setminus L(T(v'))\ne\emptyset$.
  Together with  $v'\in\child_T(u)$, these two arguments imply
  $u\preceq_T\lca_T(B)$
  which is equivalent to $Y\subseteq B_{\mathcal{H}}$.
  In summary, Condition (b) is satisfied.
\end{proof}
We are now in the position to show that r-compatibility can be decided in
  linear time.
\begin{theorem}
  \label{thm:comp-linear}
  Given a rooted tree $T$ on $X$ and a partition $\mathcal{P}$ of $X$, it can
  be decided in $O(|X|)$ time whether $\mc{P}$ and $T$ are
    r-compatible. In this case, the edge coloring $\gamma_{T,\mc{P}}$ and
  a compatible refinement can also be constructed in $O(|X|)$ time.
\end{theorem}
\begin{proof}
  We employ the sparse-table algorithm described in \cite{Bender:05},
  which, following an $O(|X|)$-preprocessing step, enables constant-time
  look up of $\lca_T(u,v)$ for any $u,v\in V(T)$. We represent
  $\gamma_{T,\mathcal{P}}$ by a (hash-based) map data structure that
  contains the $O(|X|)$ edges $e\in E(T)$ as keys, and (hash-based,
  initially-empty) sets as values, which will be filled with the elements
  in $\gamma_{T,\mathcal{P}}(e)$. The sets in
    $\mathcal{P}$ can be represented by pointers to these sets or by integer
    indices when used as colors.
  We next show that
  $\gamma_{T,\mathcal{P}}$ can be constructed in $O(|X|)$ time.  When an
  edge $e$ is colored with $A$ (i.e., $A$ is added to
  $\gamma_{T,\mathcal{P}}(e)$), we check in constant time whether $e$ still
  has at most one color. If this is not the case, we stop the algorithm
  since $\mc{P}$ and $T$ are not r-compatible
   by Prop.~\ref{prop:at-most-one-color}.
  Conversely, Prop.~\ref{prop:at-most-one-color} implies that if we color
  each edge at most once, then $\mc{P}$ and $T$ are r-compatible.

  We process every $A=\{x_1,\dots,x_k\}\in\mathcal{P}$ as follows.
  First, we initialize the set of previously visited vertices of $V(T)$ as
  $\texttt{visited}\leftarrow\emptyset$. Moreover, we initialize the
  current last common ancestor as $\texttt{curLCA}\leftarrow x_1$, which we
  will update stepwise until it equals $\lca_T(A)$ in the end. To
  this end, for each leaf $x\in\{x_2,\dots,x_k\}$ (if any), we query
  $\texttt{newLCA}=\lca_T(x,\texttt{curLCA})$ and move from $x$ upwards
  along the tree. Each edge $e=\{\parent_{T}(v),v\}$ encountered during the
  traversal is colored with $A$, and $v$ is added to \texttt{visited}.  The
  traversal stops as soon as $\parent_{T}(v)$ is in \texttt{visited} or
  equals $\texttt{newLCA}$.  In case we have
  $\texttt{curLCA}\prec_T\texttt{newLCA}$, which by definition of
  \texttt{newLCA} holds if $\texttt{curLCA}\ne\texttt{newLCA}$, we perform
  the same bottom-up traversal starting from $\texttt{curLCA}$.  As a final
  step in the processing of $x$, we set
  $\texttt{curLCA}\leftarrow\texttt{newLCA}$.  One easily verifies that,
  after processing all vertices in $A$, we have exactly colored the edges
  in the minimal subtree of $T$ that connects all leaves in $A$.
  Moreover, each edge considered in the bottom-up traversals is colored
  with $A$ and required only a constant number of constant-time queries and
  operations.  Similarly, the additional operations needed for each
  $x\in A$ (i.e., set initialization, query, comparison, and update of
  the last common ancestor) are performed in constant time.  Since the
  algorithm stops as soon as an edge would be colored with two colors, at
  most one edge is considered twice in the bottom-up traversal. Since $T$
  is a phylogenetic rooted tree, we have $|E(T)|\le 2|X|-2$. In total,
  therefore, the traversals of the tree require $O(|X|)$ operations. In
  addition, a constant effort is required for each of the $O(|X|)$ vertices
  in the disjoint sets in $\mathcal{P}$. Thus $\gamma_{T,\mathcal{P}}(e)$
  can be constructed in $O(|X|)$ time.

  It remains to show how a compatible refinement of $T$ can be
    constructed.  Put $\mathcal{H} \coloneqq \mathcal{H}(T)$ and
  $\mathfrak{Y}\coloneqq\mathfrak{Y}(\mathcal{H},\mathcal{P})$.  First note
  that all inner vertices $u$ that need to be resolved correspond to some
  $Y$ (i.e., $u=\lca_T(Y)$) such that $Y=A_{\mc{H}}$ for some
  $A\in\mathfrak{Y}$.  By Lemma~\ref{lem:Y-localColoring}, it suffices to
  solely check the colorings for the edges incident to $u$ according the
  two conditions in Lemma~\ref{lem:Y-localColoring}. Therefore, we do not
  need to consider the sets $Y=L(T(u))$ explicitly.  In this way, each edge
  in $T$ and its set of colors must be checked at most twice.  Since
  $|\gamma_{T,\mathcal{P}}(e)|\le 1$ for every edge $e\in E(T)$, this can
  be done in $O(|X|)$ time.  By Lemma~\ref{lem:HP-compatible}, $\HP$
  is a refinement of $\mathcal{H}$ that is compatible with
  $\mathcal{P}$. Instead of operating on $\mathcal{H}$ and $\mathfrak{Y}$,
  we directly construct the tree $T^*$ corresponding to $\HP$ from
  $T$ in $O(|X|)$-time as follows.  If a vertex $u=\lca_T(Y)$
  satisfies Conditions (a') and (b') in Lemma~\ref{lem:Y-localColoring},
  then the respective coloring of its edges imply that there is an
  $A\in\mathfrak{Y}$ with $A_{\mathcal{H}}=Y$ for some color $A\in\mc{P}$.
  We refine $T$ at vertex $u$ as follows: By definition, the sets
  $W_j\in\mc{H}$ whose disjoint union gives the newly-created sets $Y_{A}$
  are child clusters of $Y$ in $\HP$.  Hence, they correspond to the
  children $v_j\in\child_T(u)$ for which the edge $\{u,v_j\}$ is colored
  with $A$. Therefore, we remove all of these edges $\{u,v_j\}$, and
  instead add the edge $\{u,v_{A}\}$ and the edges $\{v_{A},v_j\}$, where
  $v_{A}$ is a newly-created vertex.  In particular, we have
  $Y_{A}=L(T(v_{A}))$.  Since this is true for all sets
  $Y_A\in\HP\setminus\mathcal{H}$, the resulting tree $T^*$ corresponds to
  the hierarchy $\HP$.  Clearly, we introduce no more than $O(|X|)$ new
  vertices. Since each edge has at most one color, at most $O(|X|)$
  operations are required.
\end{proof}

We next characterize compatibility of $\mathcal{H}$ and $\mathcal{P}$ in
terms of the edge coloring $\gamma_{T,\mathcal{P}}$ and show that
compatibility of $\mathcal{H}$ and $\mathcal{P}$ can be tested in linear time.
\begin{theorem}
  Let $\mathcal{H}$ be a hierarchy on $X$, $T$ the corresponding tree on
  $X$, and $\mathcal{P}$ a partition of $X$.  Then
  $\mathcal{H}$ and $\mathcal{P}$ are compatible if and only if there is no
  vertex $u\in V(T)$ and distinct $A,B\in\mc{P}$ such that
  $A\in\gamma_{T,\mathcal{P}}(\{u,v\})$ and
  $B\in\gamma_{T,\mathcal{P}}(\{u,v'\})$ for (not necessarily distinct)
  children $v,v'\in\child_T(u)$.
  In particular, it can be decided in $O(|X|)$ time whether $\mathcal{P}$
  and
    $T$ are compatible.
  \label{thm:compatible-iff-color}
\end{theorem}
\begin{proof}
  First suppose, for contraposition, that there is a vertex $u\in V(T)$
  with distinct $A,B\in\mc{P}$ such that $A\in\gamma_{T,\mathcal{P}}(\{u,v\})$
  and $B\in\gamma_{T,\mathcal{P}}(\{u,v'\})$ for children
  $v,v'\in\child_T(u)$.  If $v=v'$, we can apply
  Prop.~\ref{prop:at-most-one-color} to conclude that there is no refinement
  of $\mathcal{H}$ that is compatible with $\mathcal{P}$. Thus, in
  particular, $\mathcal{H}$ is not compatible with $\mathcal{P}$.  Now
  assume that $v$ and $v'$ are distinct.  We distinguish the three cases
  (a)~$u\ne\rho_T$ and
  $A,B\in\gamma_{T,\mathcal{P}}(\{\parent_{T}(u),u\})$, (b)~$u\ne\rho_T$
  and exactly one of $A$ and $B$ is in
  $\gamma_{T,\mathcal{P}}(\{\parent_{T}(u),u\})$, and (c)~$u=\rho_T$ or
  $A,B\notin\gamma_{T,\mathcal{P}}(\{\parent_{T}(u),u\})$.  In case~(a),
  $\gamma_{T,\mathcal{P}}(\{\parent_{T}(u),u\})$ contains two colors and we
  again obtain incompatibility of $\mathcal{H}$ and $\mathcal{P}$ by
  Prop.~\ref{prop:at-most-one-color}.  In case~(b), we assume w.l.o.g.\ that
  $A\notin\gamma_{T,\mathcal{P}}(\{\parent_{T}(u),u\})$ and
  $B\in\gamma_{T,\mathcal{P}}(\{\parent_{T}(u),u\})$.  Since
  $A\in\gamma_{T,\mathcal{P}}(\{u,v\})$, we have $|A|> 1$ and since
  $A\notin\gamma_{T,\mathcal{P}}(\{\parent_{T}(u),u\})$, it must hold that
  $A\subseteq L(T(u))$. The latter two arguments imply that there is a
  second edge $\{u,w\}$, $w\in \child_T(u)$ with
  $A\in\gamma_{T,\mathcal{P}}(\{u,w\})$ and, in particular, $u=\lca_T(A)$.
  Moreover, since $B\in\gamma_{T,\mathcal{P}}(\{\parent_{T}(u),u\})$, we
  have $B\cap L(T(u))\ne\emptyset$ and
  $B\setminus L(T(u))\ne\emptyset$.  In particular, therefore, $u$
  corresponds to $A_{\mathcal{H}}=L(T(u))$ and $A_{\mathcal{H}}$ is not
  the union of sets in $\mathcal{P}$.  Together with
  Thm.~\ref{thm:compatible}, this implies that $\mathcal{H}$ and
  $\mathcal{P}$ are not compatible.  In case~(c), we have, by similar
  arguments as before, that $u=\lca_T(A)=\lca_T(B)$. Hence, $u$
  corresponds to both $A_{\mathcal{H}}$ and $B_{\mathcal{H}}$ for
  distinct $A,B\in\mathcal{P}$, and thus, Condition~(ii) in
  Thm.~\ref{thm:compatible} is not satisfied. Therefore, $\mathcal{H}$ and
  $\mathcal{P}$ are not compatible.

  To prove the converse, suppose, for contraposition, that $\mc{H}$
  and $\mc{P}$ are not compatible. Hence, Condition~(i) or~(ii) in
  Thm.~\ref{thm:compatible} is not satisfied.  If Condition~(i) is not
  satisfied, then there is some $A\in\mc{H}$ such that
  $A_{\mc{H}}$ is not the union of sets in $\mc{P}$.  Since
  $A\subseteq A_{\mc{H}}$ by definition, the latter implies that
  there must be some $B\in\mc{P}\setminus\{A\}$ such that
  $B\cap A_{\mc{H}}\ne\emptyset$ and
  $B\setminus A_{\mc{H}}\ne\emptyset$.  Moreover, since $A$ and
  $B$ are disjoint and non-empty and both contain elements that are in
  $A_{\mc{H}}$, the vertex $u\coloneqq \lca_T(A)$ corresponding to
  $A_{\mc{H}}$ is an inner vertex and has (not necessarily distinct)
  children $v,v'\in\child_T(u)$ such that
  $A\in\gamma_{T,\mc{P}}(\{u,v\})$ and
  $B\in\gamma_{T,\mc{P}}(\{u,v'\})$.  If Condition~(ii) is not
  satisfied, i.e., $A_{\mathcal{H}}=B_{\mc{H}}$ for two distinct
  $A,B\in\mc{P}$, we can apply similar arguments to conclude that
  $u\coloneqq\lca_T(A)=\lca_T(B)$ has children $v,v'\in\child_T(u)$
  such that $A\in\gamma_{T,\mc{P}}(\{u,v\})$ and
  $B\in\gamma_{T,\mc{P}}(\{u,v'\})$.

  It remains to show that compatibility of $\mc{P}$ and $T$ can be decided
    in $O(|X|)$. Compatibility implies r-compatibility which can be checked in
    $O(|X|)$ by Thm.~\ref{thm:comp-linear}. In particular, the edge coloring
    $\gamma_{T,\mathcal{P}}$ can be constructed with the same complexity in
    this case.
  The condition whether or not there is a vertex $u\in V(T)$ and distinct
    $A,B\in\mc{P}$ such that $A\in\gamma_{T,\mathcal{P}}(\{u,v\})$ and
    $B\in\gamma_{T,\mathcal{P}}(\{u,v'\})$ for (not necessarily distinct)
    children $v,v'\in\child_T(u)$
  can easily be checked in $O(|X|)$ time by counting, for each vertex in an
  arbitrary traversal, the number of colors appearing on the edges leading to
  its children.
\end{proof}

Similar to Lemma~\ref{lem:min-size}, we obtain here a result for
maximum-sized sets of separating edges.  Since
any edge $e\in E(T)$ with $\gamma_{T,\mathcal{P}}(e)\neq \emptyset$
cannot be a separating edge, and any edge $e$ for which
  $\gamma_{T,\mathcal{P}}(e)=\emptyset$ can always be added as a separating
  edge, we obtain
\begin{corollary}
  Suppose that $\mathcal{P}$ and $T$ are compatible. Then, there is a
  unique maximum-sized set of separating edges $H^*$, which is given by the set
  of
  edges $e\in E(T)$ for which $\gamma_{T,\mathcal{P}}(e)=\emptyset$.
  This maximum-sized set $H^*$ can be computed in $O(|X|)$ time.
  \label{cor:MaxHGT}
\end{corollary}
For a minimum-sized set of separating edges $H^*$ for compatible $\mathcal{P}$
and
$T$, the cardinality of $H^*$ can be expressed as a function of $|\mathcal{P}|$
alone (cf.\ Lemma~\ref{lem:min-size}), and is thus independent of $T$.
However, the latter is not the case for a maximum-sized $H^*$ set of separating
edges.
To see this, consider a partition $\mathcal{P}$ of $X$ consisting of all
singletons.
Clearly, we have $H^*=E(T)$ for any tree on $X$ where $|E(T)|$ varies depending
on how resolved a specific tree $T$ is.

\smallskip

In what follows, we investigate the complexity of the problem of
recognizing compatibility of systems
$\mathfrak{P}=\{\mathcal{P}_1,\mathcal{P}_2,\dots,\mathcal{P}_k\}$ of
partitions of $X$ with (refinements of) trees.
In Section~\ref{sec:overview}, we have introduced the two
closely related problems asking whether a tree $T$ admits a refinement that is
compatible with all partitions in $\mathfrak{P}$, \textsc{CompaTP}, and whether
a compatible tree exists at all, \textsc{ExistTP}.
To this end, we first show that  \textsc{ExistTP}
is a simple translation of the \textsc{Symm-Fitch Recognition} problem, which is
NP-complete \cite[Thm.4.2]{Hellmuth:21x}.
In particular, this discussion will yield NP-completeness of
\textsc{CompaTP} and \textsc{ExistTP} in Thm.~\ref{thm:NPc} below.

The concept of compatible partitions and trees is intimately related to
so-called undirected Fitch graphs $G$, that is, complete multipartite
graphs whose maximal independent sets form a partition $\mathcal{P}$ of
$V(G)$ \cite{Hellmuth:18a}.  To be more precise, for a given tree $T$ with
leaf set $X$ and subset $H\subseteq E(T)$, an undirected Fitch graph
$G = (X,E)$ has an edge $\{x,y\}\in E$ if and only if there is an edge in
$H$ that lies on the path between $x$ and $y$ in $T$. Therefore,
$\{x,y\}\notin E$ if and only if $x$ and $y$ are contained in the same set
$B\in\mathcal{P}\coloneqq \mathcal{F}(T,H)$.  This construction was
generalized in \cite{Hel:19,Hellmuth:20c} to \emph{Fitch maps} that allow
multiple colors. In the following paragraph we briefly summarize the
construction of \emph{symmetrized Fitch maps} \cite{Hellmuth:21x}. We then
show that a symmetrized Fitch map $\eps$ can be interpreted as a partition
system $\mathfrak{P}$ on $X$ that is compatible with a suitably chosen
tree $T$.

Let $M\coloneqq\{1,\dots,k\}$ be a set of colors for some
$k\in\mathbb{N}$. Moreover,
for a set $X$, we write
$\irr{X\times X}\coloneqq (X\times X) \setminus \{(x,x)\mid x\in X\}$.  An
\emph{edge-colored tree $(T,\lambda)$} is a tree $T$ together with a map
$\lambda: E(T) \to 2^{M}$. Note that $\lambda$ can be chosen
  arbitrarily in contrast to the $\mathcal{P}$-coloring
  $\gamma_{T,\mathcal{P}}$ of $T$ as in Def.\ \ref{def:P-col}. An edge
  $e\in E(T)$ is an $m$-edge if $m\in \lambda(e)$ for some $m\in M$.

A map $\eXM$ is a \emph{symmetrized Fitch map}
if there is an edge-colored tree $(T,\lambda)$ with leaf set $X$ and edge
coloring $\lambda: E(T)\to 2^{M}$ such that for every pair
$(x,y)\in \irr{X\times X}$ it holds that
\begin{equation*}
  m \in \eps(x,y) \iff \textnormal{ there is an }
  m\textnormal{-edge on the path from } x \textnormal{ to } y.
\end{equation*}
In this case, we say that $\eXM$ \emph{is explained by} $(T,\lambda)$.

For an arbitrary map $\eXM$ and each $m\in M$, the \emph{monochromatic map
  (induced by $m$)} is given by
$\eps_m(x,y) \coloneqq \eps(x,y)\setminus (M\setminus \{m\})$.  By
definition, we have $\eps_m(x,y)\in \{\emptyset, \{m\}\}$ and, in
particular, $\eps_m(x,y) = \{m\}$ if and only if $m\in \eps(x,y)$.  If
moreover $\eps$ (and thus $\eps_m$) is a symmetrized Fitch map, then there
is a tree $(T,\lambda)$ such that $\eps_m(x,y) = \{m\}$ if and only if
there is an $m$-edge on the path from $x$ to $y$.  In \cite{Hellmuth:18a},
it was shown that the graph representations $G_m=(X,E_m)$ of
(monochromatic) symmetrized Fitch relations, given by $\{x,y\}\in E$ if and
only if $\eps_m(x,y) = \{m\}$, coincide with the class of complete
multipartite graphs. Therefore, they can uniquely be represented by a
partition $\mathcal{P}_m^{\eps}$ of $X$ in a way that each set
$A\in\mathcal{P}_m^{\eps}$ corresponds to a maximal independent subset of
$X$, i.e.\ $\{x,y\}\notin E_m$ for all $x,y\in A$. In particular, it holds
that $x,y\in X$ are elements of distinct sets of
$\mathcal{P}_m^{\eps}$ if and only
if $m\in \eps(x,y)$. Thus, if all monochromatic maps are
symmetrized Fitch maps,
$\mathfrak{P}^{\eps}\coloneqq\{ \mathcal{P}_m^{\eps} \mid m\in M\}$ is a
partition system on $X$.

\begin{lemma}
  \label{lem:symfitch-existtp-equiv}
  Let $\eXM$ be a map such that the monochromatic maps $\eps_m$ are
  symmetrized Fitch maps for all $m\in M$, and $\overline T$ be an unrooted
  tree with leaf set $X$.  Then, there is an edge-coloring $\lambda$ such
  that $(\overline T,\lambda)$ explains $\eps$ if and only if the partition
  $\mathcal{P}_m^{\eps}$ of $X$ is compatible with $\overline T$ for all
  $m\in M$.
\end{lemma}
\begin{proof}
  First note that, since the monochromatic maps $\eps_m$ are symmetrized
  Fitch maps for all $m\in M$, the partitions $\mathcal{P}_m^{\eps}$ of $X$
  are all well-defined.  For the case $|X|\in \{1,2\}$, the statement is
  trivially true since, in this case, every of the (at most two)
    possible partitions of $X$ are compatible with a unique tree on $X$.
  Thus, assume that $|X|\ge 3$. Let $T$ be any rooted version of
  $\overline T$.  Since $|X|\ge 3$, we can apply Prop.~\ref{prop:S-root},
  to conclude that $\overline T$ is compatible with some partition
  $\mathcal{P}$ of $X$ if and only if $T$ is compatible with
  $\mathcal{P}$. Hence, it suffices to show the statements for $T$.

  Assume there is an edge-coloring $\lambda$ such that $(T,\lambda)$
  explains $\eps$.  Put $H_m \coloneqq \{e\in E(T)\mid m\in \lambda(e)\}$,
  $m\in M$.  By construction, we have for all distinct $x,y\in X$ that
  \begin{enumerate}[nolistsep]
    \item[] $x$ and $y$ are in distinct sets of $\mathcal{P}_m^{\eps}$
    \item[] $\iff$ $m\in\eps(x,y)$
    \item[] $\iff$ $m\in \lambda(e)$ for some $e\in E(T)$ on the path
    connecting
    $x$ and $y$
    \item[] $\iff$ there is an edge in $H_m$ on the path connecting
    $x$ and $y$
    \item[] $\iff$ $x$ and $y$ are in distinct sets of $\mathcal{F}(T,H_m)$.
  \end{enumerate}
  Hence, $\mathcal{P}_m^{\eps}=\mathcal{F}(T,H_m)$ and
  $\mathcal{P}_m^{\eps}$ is compatible with $T$ for all $m\in M$.

  Now assume that the partition $\mathcal{P}_m^{\eps}$ of $X$ is compatible
  with $T$ for all $m\in M$. For each $m\in M$, define
  $\lambda_m(e) = \{m\}$ for all edges
  $e = \{\parent(\lca_T(B)),\lca_T(B)\}$ for some
  $B\in \mathcal{P}_m^{\eps}$ with $\lca_T(B)\neq \rho_T$ and, for all
  remaining edges $e$ put $\lambda_m(e) = \emptyset$.  By construction, we
  have for $e\in E(T)$ that $\lambda_m(e) = \{m\}$ if and only if $e$ is
  contained the set $H_m$ as specified in Eq.~\eqref{eq:hgt-edges}.  By
  Cor.~\ref{cor:set-H}, we have
  $\mathcal{P}_m^{\eps} = \mathcal{F}(T,H_m)$.  Hence, $m\in \eps(x,y)$ if
  and only if there is an edge $e$ along the path between $x$ and $y$ with
  $\lambda_m(e) = \{m\}$.  Now set
  $\lambda(e) = \cup_{m\in M} \lambda_m(e)$ for all $e\in T$ to obtain the
  final coloring such that $(T,\lambda)$ explains $\eps$.
\end{proof}

\begin{lemma}
  \textsc{Symm-Fitch Recognition} remains NP-hard if the monochromatic maps
  $\eps_m$ are symmetrized Fitch maps for all $m\in M$.
  \label{lem:NP-c-all-monochromatic}
\end{lemma}
\begin{proof}
  As shown in \cite[Thm.~4.2]{Hellmuth:21x}, \textsc{Symm-Fitch
  Recognition} is NP-complete.
  Note, if for a map $\eXM$ there is an $m\in M$ such that $\eps_m$ is not
  a monochromatic Fitch match, then $\eps$ cannot be a symmetrized Fitch
  map; a property that can be checked in polynomial-time for all $m\in M$
  \cite{Hellmuth:18a}.  Hence, under the assumption that $P\neq NP$, the
  NP-hard instances must, in particular, be included within the instances
  of \textsc{Symm-Fitch Recognition} for which $\eps_m$ is a monochromatic
  Fitch match for all $m\in M$.
\end{proof}

We are now in the position to establish NP-completeness of \textsc{ExistTP}
and \textsc{CompaTP}.

\begin{theorem}\label{thm:NPc}
  \textsc{ExistTP} and \textsc{CompaTP} are NP-complete.
\end{theorem}
\begin{proof}
  By Thm.~\ref{thm:compatible-iff-color} and
  the fact that one can test in polynomial time whether $T^*$ is a
  refinement of $T$ by comparing their hierarchies, \textsc{ExistTP}
  and \textsc{CompaTP} are contained in the class NP.
  By Lemma~\ref{lem:symfitch-existtp-equiv},
  \textsc{ExistTP} is equivalent to the \textsc{Symm-Fitch
    Recognition} problem restricted to a certain set of instances for which,
    by Lemma~\ref{lem:NP-c-all-monochromatic},
  the problem remains NP-hard.
    Asking for a
    tree that is compatible with $\mathfrak{P}$ is equivalent to asking
    whether there exists a refinement $T^*$ of the star tree that is
    compatible with $\mathfrak{P}$. Therefore, \textsc{ExistTP} is a
    special instance of \textsc{CompaTP} and NP-hardness of
    \textsc{ExistTP} implies that \textsc{CompaTP} is also NP-hard. Since
    both problems are in class NP, \textsc{ExistTP} and \textsc{CompaTP}
    are NP-complete.
\end{proof}

The latter problems become fixed-parameter tractable
and thus, easier, if
$T$ is ``almost binary''. In \cite{Schaller:21c}, the resolution of a rooted
tree is quantified by the normalized parameter
$\res(T)\coloneqq (|V|-|X|-1)/(|X|-2)$, which varies between $0$ (star
tree) and $1$ (binary tree). The quantity
$h(T)\coloneqq 2|X|-|E(T)|-2 = (|X|-2)(1-\res(T))$ correspondingly measures
how much $T$ deviates from being binary.  Now let $V^*$ be the set of
non-binary inner vertices of $T$.  Writing $h_v\coloneqq |\child_T(v)|-2$
for the number of ``excess children'' at the inner vertex $v\in V^*$, one
easily checks that $h(T) = \sum_{v\in V^*} h_v$.  Now suppose $h(T)\le
h$. Then all possible binary refinements of $T$ are obtained by inserting
an arbitrary binary tree between each non-binary vertex $v$ of $T$ and its
children $\child(v)$. It is well known the that the number of binary rooted
leaf-labeled trees on $d_v=|\child(v)|=h_v+2$ leaves is
$(2d_v - 3)!!=(2h_v+1)!!$ \cite{Felsenstein:78}. The total number of binary
refinements is therefore $\prod_{v\in V^*} (2h_v+1)!!$. From the definition
of the double factorial $(2n+1)!! = 1\cdot 3\cdot\dots...\cdot (2n+1)$ we
see that, after omitting the leading factors $1$, $(2h_v+1)!!$ has exactly
$h_v$ factors for each $v\in V^*$, and thus, $\prod_{v\in V^*} (2h_v+1)!!$
has exactly $\sum_{v\in V^*} h_v$ factors. Similarly, $(2h+1)!!$ has $h$
contributing factors greater than $1$. By ordering these factors in
$\prod_{v\in V^*} (2h_v+1)!!$ and $(2h+1)!!$, resp., one easily verifies
that, since $\sum_{v\in V^*} h_v\le h$, the second product has at least as
many factors as the first, and moreover, each of them is not smaller than
the corresponding factor (w.r.t.\ the ordering) in the first product (if
existent). Note that equality holds if and only if $V^*$ comprises a
single vertex with $h+2$ children.
Each binary refinement can be checked in $O(|\mathfrak{P}|\,|X|)$ time for
consistency with $\mathfrak{P}$ since the consistency check for a single
partition can be performed in $O(|X|)$ time by Thm.~\ref{thm:comp-linear}
below. Thus there is an $O((2h+1)!!\,|\mathfrak{P}|\,|X|)$ algorithm and
thus, \textsc{CompaTP}   is FPT for the parameter $h$.

\section{Concluding Remarks}

We have characterized the compatibility of a partition $\mathcal{P}$ with a
hierarchy $\mathcal{H}$. The concept of compatibility considered here is
much more general than that of a ``representative partition'', i.e., the
cutting of hierarchy $\mathcal{H}$ at a particular aggregation
level. Instead, it amounts to disconnecting the corresponding tree $T$ at
an arbitrary set of edges $H\subseteq E(T)$, i.e., at a subset of the
splits $\mathfrak{S}(T)$. In Section~\ref{sect:refine}, we have
characterized when a refinement of $T^*$ of a tree $T$ exists such that
$\mathcal{P}$ and $T^*$ are compatible. For practical application, it may
be relevant to allow more general operations on the tree $T$. A natural
generalization is to allow not only refinements but also edge contraction
while editing $T$ into a tree $T'$ that is compatible with a partition
$\mathcal{P}$ of interest. This amounts to minimizing the cardinality
$|\mathfrak{S}(\overline{T})\symdiff\mathfrak{S}(\overline{T'})|$ of the
symmetric difference of the corresponding split systems, i.e., the
Robinson-Foulds distance of $T$ and $T'$.

In Section~\ref{sec:splits}, we have considered tree-like split
systems, i.e., split systems that can be represented by unrooted
  trees.  Thm.~\ref{thm:splitcomp} and Cor.~\ref{cor:splitrefcomp},
however, suggest to consider the compatibility of a partition $\mathcal{P}$
and a split system $\mathfrak{S}$ in a more general setting.  These
characterizations may then serve as convenient definitions in a more
general context: We may say that a partition $\mathcal{P}$ and a split
system $\mathfrak{S}$ are compatible if there is a set of splits
$\mathfrak{H}$ such that $\bigwedge\mathfrak{H}=\mathcal{P}$ and
$\mathfrak{H}\subseteq\mathfrak{S}$. In order to handle refinements in this
setting, one would ask whether there is a set of splits $\mathfrak{H}$ such
that $\bigwedge\mathfrak{H}=\mathcal{P}$. Without further restrictions,
$\mathfrak{H}=\mathfrak{S}_{\mathcal{P}}$ always provides a positive answer
to this question.  In analogy with our discussion above, it therefore seems
natural to consider only split systems $\mathfrak{S}$ that belong to a
certain class of interest. A refinement will be feasible only if the split
system $\mathfrak{H}\cup\mathfrak{S}$ again belongs to the desired class.
Natural generalizations of tree-like split systems to which the notion
  of compatibility may be applied include circular and weakly compatible
split systems \cite{Semple:03,Dress:12}, or the even more general
\emph{Teutoburgan} split systems considered in \cite{Huber:06}. The
suggested definition of compatibility in terms of split systems also
provides the natural generalization to the framework of X-trees
\cite{Semple:03,Dress:12}, i.e., to trees in which the set of taxa $X$ is
not restricted to the leaves of $T$ but may also appear as inner vertices
of $T$. This amounts to lifting our requirement that the trivial splits
$\{x\}|(X\setminus\{x\})$ must be included in $\mathfrak{S}(\overline{T})$.

We have seen in Section~\ref{sec:compl-algos} that compatibility of
$\mathcal{P}$ and $\mathcal{H}$ and the existence of a refinement
$\mathcal{H}^*$ can be decided in linear time, while the extension to
arbitrary partition systems $\mathfrak{P}$ is NP-complete. Several
interesting open questions remain concerning the computational complexity
of the \textsc{Compatibility of Tree and Partition System} problem and
  the \textsc{Existence of Tree compatible with Partition System}
  problem. Do these problems remain NP-complete if the tree corresponding
to $\mathcal{H}$ has bounded degree? What if the number $|\mathfrak{P}|$ of
input partitions is kept constant?  Since the related \textsc{Symm-Fitch
  Recognition} problem \cite{Hellmuth:21x} is in turn closely related to
the problem of \textsc{Unrooted Tree Compatibility} \cite{Bryant:06}, which
is known to be FPT in number of input trees, it is not unlikely that
\textsc{Compatibility of Tree and Partition System} is FPT in the number of
partitions. Furthermore, it is interesting to ask whether there are (easily
recognizable) subclasses of partition systems for which
\textsc{CompaTP} and \textsc{ExistTP} become tractable. Interesting
candidates are the braids of partitions appearing in image analysis
\cite{Kiran:15,Tochon:19}, or the hierarchical partition systems considered
in \cite{Huber:14}, for which
$\bigcup_{x\in X}\{\{x\}\}\cup
\bigcup_{\mathcal{P}\in\mathfrak{P}}\mathcal{P} \cup\{X\}$ forms a
hierarchy.

\subsection*{Acknowledgments}

This work was funded in part by the \emph{Deutsche Forschungsgemeinschaft}. 
We  thank the anonymous referees for the constructive comments and 
recommendations which helped to significantly improve the readability and 
quality of the paper. 

\bibliography{preprint2-HandP}

\end{document}